\documentclass[a4paper,11pt]{article}
\usepackage{jheppub} 
\usepackage{lineno}
\usepackage{graphicx}
\usepackage{amssymb,amsmath}
\usepackage{bm}
\usepackage{dcolumn}
\usepackage{subfigure}
\usepackage{float}
\usepackage[OT1]{fontenc} 
\usepackage{url}
\usepackage{slashed}
\usepackage{color}
\usepackage{verbatim}
\usepackage{enumitem}
\usepackage{txfonts}
\usepackage{soul,physics}
\usepackage{mathrsfs}

\newcommand{\Bb}{\mathscr{B}}
\newcommand{\nothing}[1]{}
\renewcommand{\t}{\tau}
\hypersetup{pdfpagemode=FullScreen,colorlinks=true,breaklinks,urlcolor=blue,linkcolor=blue,citecolor=blue}
\usepackage{natbib}

\usepackage{subfigure}
\usepackage{comment}
\usepackage{hyperref}
\usepackage{tikz,fp}
\usepackage{tikz-cd}
\usetikzlibrary{arrows}
\usetikzlibrary{intersections}
\usetikzlibrary{shapes.geometric}
\usetikzlibrary{decorations.pathmorphing, patterns,shapes,fixedpointarithmetic}
\usetikzlibrary{decorations.markings}

\newcommand{\boxalign}[2][0.97\textwidth]{
 \par\noindent\tikzstyle{mybox} = [draw=black,inner sep=6pt]
 \begin{center}\begin{tikzpicture}
  \node [mybox] (box){%
   \begin{minipage}{#1}{\vspace{-5mm}#2}\end{minipage}
  };
 \end{tikzpicture}\end{center}
}

\pgfdeclarepatternformonly{south west lines}{\pgfqpoint{-0pt}{-0pt}}{\pgfqpoint{3pt}{3pt}}{\pgfqpoint{3pt}{3pt}}{
	\pgfsetlinewidth{0.4pt}
	\pgfpathmoveto{\pgfqpoint{0pt}{0pt}}
	\pgfpathlineto{\pgfqpoint{3pt}{3pt}}
	\pgfpathmoveto{\pgfqpoint{2.8pt}{-.2pt}}
	\pgfpathlineto{\pgfqpoint{3.2pt}{.2pt}}
	\pgfpathmoveto{\pgfqpoint{-.2pt}{2.8pt}}
	\pgfpathlineto{\pgfqpoint{.2pt}{3.2pt}}
	\pgfusepath{stroke}}

\tikzset{
	mid arrow/.style={postaction={decorate,decoration={
				markings,
				mark=at position .575 with {\arrow{stealth}}
	}}},
	near arrow/.style={postaction={decorate,decoration={
				markings,
				mark=at position .275 with {\arrow{stealth}}
	}}},
	far arrow/.style={postaction={decorate,decoration={
				markings,
				mark=at position .800 with {\arrow{stealth}}
	}}},
	snake arrow/.style={fixed point arithmetic, decorate, decoration={snake,amplitude=2pt, segment length=11pt},postaction={decoration={markings,mark=at position 0.625 with {\arrow{stealth}}},decorate}},
}
\tikzset{
  baseline = -0.5ex,
  wavy/.style = {
    thick,
    decorate,
    decoration={snake,amplitude=2pt,segment length=5pt}},
  sdot/.style = {
    circle,
    draw=none,
    fill=black,
    minimum size=2.5pt,
    inner sep=0pt},
  bdot/.style = {
    circle,
    draw=none,
    fill=black,
    minimum size=4pt,
    inner sep=0pt},
  svertex/.style = {
    circle,
    draw=black,
    thick,
    fill=lightgray,
    minimum size=8pt,
    inner sep=1pt},
  bvertex/.style = {
    circle,
    draw=black,
    thick,
    fill=lightgray,
    minimum size=24pt},
  bvertexsmall/.style = {
    circle,
    draw=black,
    thick,
    fill=lightgray,
    minimum size=7pt},
  bvertexnormal/.style = {
    circle,
    draw=black,
    thick,
    fill=lightgray,
    minimum size=16pt},
  dvertex/.style = {
    circle,
    draw=black,
    thick,
    fill=gray,
    minimum size=25pt}}

\newcommand{\iu}{{i\mkern1mu}}
\newcommand*\diff{\mathop{}\!\mathrm{d}}

\newcommand{\rel}{t_{\textrm{rel}}}

\DeclareMathOperator{\sgn}{sgn}

\makeatletter
\newcommand*{\rom}[1]{\expandafter\@slowromancap\romannumeral #1@}
\makeatother

\title{The Magnetic Maze: A System With Tunable Scale Invariance}

\author[a,b]{Tian-Gang Zhou,}
\author[c,d]{Michael Winer,}
\author[b]{and Brian Swingle}

\affiliation[a]{Institute for Advanced Study, Tsinghua University, Beijing, 100084, China}
\affiliation[b]{Department of Physics, Brandeis University, Waltham, Massachusetts 02453, USA}
\affiliation[c]{Department of Physics, University of Maryland, College Park, Maryland 20741, USA}
\affiliation[d]{
Institute for Advanced Study, Princeton, NJ 08540, USA}

\abstract{
Random magnetic field configurations are ubiquitous in nature. Such fields lead to a variety of dynamical phenomena, including localization and glassy physics in some condensed matter systems and novel transport processes in astrophysical systems. Here we consider the physics of a charged quantum particle moving in a ``magnetic maze'': a high-dimensional space filled with a randomly chosen vector potential and a corresponding magnetic field. We derive a path integral description of the model by introducing appropriate collective variables and integrating out the random vector potential, and we solve for the dynamics in the limit of large dimensionality. We derive and analyze the equations of motion for Euclidean and real-time dynamics, and we calculate out-of-time-order correlators. We show that a special choice of vector potential correlations gives rise, in the low temperature limit, to a novel scale-invariant quantum theory with a tunable dynamical exponent. Moreover, we show that the theory is chaotic with a tunable chaos exponent which approaches the chaos bound at low temperature and strong coupling.
}

\begin{document}
\maketitle
\flushbottom

\section{Introduction}

The fate of randomness in physical systems has long been an intriguing topic. There are many well-known examples that consider quantum particles moving in random scalar or vector potentials. Such randomness often leads to non-ergodic dynamics, manifesting in behaviors like glassiness and localization. One classic model is the $p$-spherical model \cite{Crisanti1992,Crisanti1993Spherical,Crisanti1995ThoulessAndersonPalmerAT,Cugliandolo2001,Cugliandolo1993Analytical,Anous:2021eqj}, in which a quantum particle moves in high dimensions in a random scalar potential. This is a standard solvable model of glassy physics with quenched disorder. Another related model is Parisi's hypercube model \cite{Parisi_1994,PhysRevLett.132.081601,Jia_2020,Marinari_1995}, in which a particle jumps among the vertices of a high-dimensional hypercube with random magnetic flux through each face. The hypercube model was motivated by random Josephson junctions. 
Still another class of models is motivated by various two-dimensional devices with magnetic impurities that exhibit spatially random vector potentials. These models typically result in electron localization physics \cite{Shoucheng2dRandomMagneticFlux1994,Shoucheng2dRandomMagneticFlux1995}, although these two-dimensional results are not easily generalized to higher dimensions. 

However, certain models can evade non-ergodic behavior and exhibit novel forms of low-energy dynamics. A well-known example is the Sachdev-Ye model, along with its extension, the Sachdev-Ye-Kitaev model \cite{Maldacena_2016,chen2020remarksreplicamethodsachdevyekitaev,Rosenhaus_2019,sachdev2024quantumspinglassessachdevyekitaev,Sachdev_1993,hartnoll2018holographicquantummatter}. Recently, a spin-based version of the SYK model was studied as part of a search for bosonic models with possible SYK-like behavior \cite{swingle2023bosonicmodelquantumholography,Hanada:2023rkf}. Although there is ample evidence that these systems can avoid a glass-like phase with replica symmetry breaking at low temperature, we wanted a system that would be manifestly replica symmetric. We then wanted to understand the potential for exotic quantum dynamics that might emerge at low energy.

\begin{figure}[htb]
    \centering
    \includegraphics[width=0.75\linewidth]{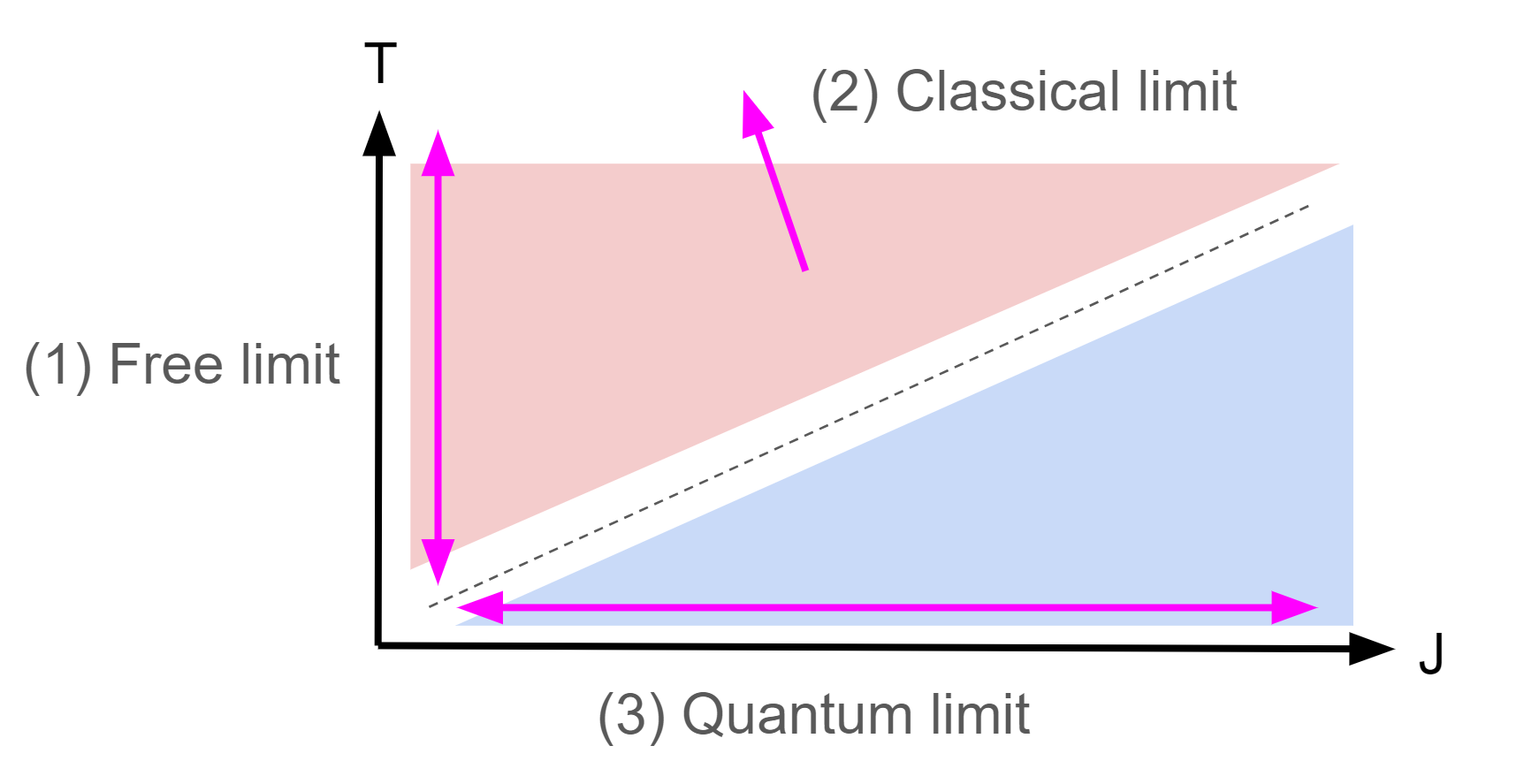}
    \caption{A rough schematic of the phase diagram of the magnetic maze as a function of temperature ($T = 1/\beta$) and coupling ($J$). There is only one equilibrium phase, but there several distinct regimes worthy of note. The diagonal dashed line denotes a cross-over from a high-temperature classical regime (red) to a low-temperature quantum regime (blue). We note three solvable limits: (1) the free limit, (2) the classical limit, and (3) the quantum limit. For (1), we simply have a free quantum particle in $N$ dimensions. For (2), the thermodynamics is that of a particle with quadratic dispersion, $z=2$, and the dynamics is chaotic with the particle moving diffusively at the longest times. For (3), the thermodynamics is that of a particle with dispersion set by a tunable dynamical exponent, $z(J)$, and the zero-temperature theory is scale invariant. At non-zero but low temperature, the dynamics remain chaotic with a relaxation time $t_{\text{rel}} = c_1(J) \beta$, a diffusivity $D = c_2(J) \beta^{2/z(J)-1}$, and a chaos exponent $\kappa = c_3(J) \frac{2\pi}{\beta}$. The quoted results are all for the special choice of vector potential correlations in Eq.~\eqref{eq:interesting_f}. }
    \label{fig:mm_phase_diagram}
\end{figure}

Hence, in this paper we study a system---the ``magnetic maze'' (MM)---consisting of a single non-relativistic charged quantum particle moving in many spatial dimensions under the influence of a random magnetic vector potential. The vector potential is time-independent but varies randomly in space with translation-invariant spatial correlations. Focusing on a special case of power-law decaying vector potential correlations, we study the thermodynamics, real-time dynamics, and Lyapunov physics as a function of temperature and coupling. The large spatial dimensionality, $N$, provides a control parameter for our analysis. Our findings are summarized as follows; see also Figure~\ref{fig:mm_phase_diagram}.

At high temperature, the system is effectively classical, and we do not expect the physics to depend sensitively on the precise choice of vector potential correlations. We show that its thermodynamics are those of a free particle and that it exhibits chaotic dynamics in which the large-scale motion of the particle is diffusive, velocity auto-correlations decay exponentially in time, and there is a positive Lyapunov exponent\footnote{These results are obtained primarily for a specific choice of vector potential correlations discussed below. We consider a more general class of correlations in Appendix~\ref{app:other_f}, primarily in the Euclidean context. Within this more general class, we have also studied a few examples of real-time dynamics in which we obtained the same signatures of chaotic dynamics.}.

At low temperature, quantum effects manifest and the physical properties depend sensitively on the choice of vector potential correlations. We show that, for a special choice of power-law spatial correlations, the system develops an emergent scaling symmetry at low energy. Whereas the high-temperature thermodynamics is identical to that of a particle with energy, $E$, related to momentum, $p$, by the usual non-relativistic dispersion, $E \sim p^2$, the low-temperature thermodynamics is instead determined by an effective dispersion, $E \sim p^z$ with a dynamical exponent $z$ which is tunable via the strength of the magnetic field, reminiscent of \cite{Haldar_2018}.

The low temperature dynamics is also chaotic, and for the same special choice of power-law spatial correlations, the emergent scaling symmetry determines the dynamical properties. In particular, there is a single timescale $\beta = \frac{k_B T}{\hbar}$ and a single length scale $\ell_\beta \sim \beta^{1/z}$ which set the dynamical properties. The decay rate of the velocity auto-correlation function and the Lyapunov exponent determined from an out-of-time-order correlator are both of order $ \beta^{-1}$, and the diffusivity is of order $D\sim \ell_\beta^2/\beta \sim \beta^{2/z-1}$. Henceforth, we set $k_B = \hbar = 1$; Appendix~\ref{app:crash_course} discusses the physical scales in the MM.

For other choices of vector potential correlations, the low-energy physics changes significantly. For example, when the spatial correlations decay more rapidly than the special case with scaling symmetry, the low temperature properties are still chaotic but more closely resemble those of a weakly perturbed free theory, with a relaxation rate and chaos exponent that vanish more rapidly than $\beta^{-1}$ at large $\beta$. We investigate this case and the opposite case of uniform magnetic field (corresponding to very long-ranged correlations) in Appendices \ref{app:crash_course} (uniform field) and \ref{app:other_f} (more rapid decay of correlations).

These results are obtained via large $N$ path integral methods \cite{HOOFT1974461,Maldacena_2016}. We formulate the system on a thermal contour to access thermodynamics, on a Schwinger-Keldysh contour \cite{Haehl_2017,CHOU19851,kamenev_2011,Kamenev_2009,Keldysh:1964ud} to access dynamics, and on a contour with two time-folds to access out-of-time-order correlations and Lyapunov physics. Because we use large $N$ as a control parameter, our results correspond to taking the limit $N \to \infty$ first before the low temperature limit. It is, of course, interesting to study $1/N$ corrections and the finite dimensional fate of the phenomena we find, but we do not pursue that here.

Central to our study is the result, discussed below, that the MM cannot have static correlations between different replicas and thus cannot have an equilibrium glassy phase. The dynamics which emerges at low energy is instead that of a scale-invariant chaotic quantum system, one which features a tunable dynamical exponent. Moreover, at the lowest temperature and largest couplings that we were able to access (figure \ref{fig:Lyapunov_lowT}), we found a chaos exponent that comes within $15 \%$ of the MSS bound \cite{MSS, Shenker_2014,Blake_2021}. We have not been able to analytically show that the model becomes maximally chaotic at infinite coupling, but we see no evidence of sub-maximal saturation in the chaos exponent. Hence, this system may provide a new example of maximal chaos.

In the remainder of the introduction, we will introduce the model in more detail and give a guide to the structure of the paper. We will also throughout the paper compare and contrast our findings with the properties of the Sachdev-Ye-Kitaev model. 

\subsection{Setup and Overview}


The Magnetic Maze consists of a single particle in $N$-dimensional space subject to a random vector potential $A_i(x)$. In order to keep our partition functions finite, we will sometimes consider a confining potential $V_{\text{conf}}=\frac{\epsilon}{2}\sum_i x_i^2$, but most of our results are most interesting when $\epsilon$ is far smaller than any other scale. The Hamiltonian is thus
\begin{equation}
    H=\sum_{i=1}^N \left[ \frac {1}{2m}(p_i-A_i(x))^2+\frac{\epsilon}{2}x_i^2 \right].
    \label{eq:bigHam}
\end{equation}
The elements of $A$ are picked from a Gaussian distribution with mean zero. The covariance of $A$ falls off as $\overline{A_i(x)A_j(y)}=f(|x-y|^2/N) \delta_{ij}$. We are particularly interested in the case where $f(x^2/N)\sim \frac 1{x^2/N}$ for large $x$, which is the special power-law form highlighted in the introduction. We take the function to be
\begin{equation}
    f(x^2/N)=\frac{J^2}{\ell^2+x^2/N},
    \label{eq:interesting_f}
\end{equation} 
where $\ell$ is some short-distance scale and $J$ is a dimensionless measure of the field strength. 

Although our focus is \eqref{eq:interesting_f}, we note that the model makes sense for other choices of $f$. We address some of these in Appendices: the case of a uniform field (Appendix \ref{app:crash_course}), which is an integrable system with $f(x^2/N)=C-\frac 12 \Bb^2x^2/N$, and the case of power-law correlations with different power-law exponent at large $x$  (Appendix \ref{app:other_f}). As we show, \eqref{eq:interesting_f} is the most interesting choice because it leads a host of interesting zero- and low-temperature properties including emergent scaling symmetry and near-maximal chaos. By contrast, the high temperature properties are less sensitive to the form of $f$.

This paper focuses on studying the classical and quantum behavior of \eqref{eq:bigHam} with $f$ given by \eqref{eq:interesting_f}. Our main quantity of interest will be $\Delta(t_1,t_2)=\frac 1N \sum_i\mathcal T (x_i(t_1)-x_i(t_2))^2$, which can be interpreted as the squared distance traveled between $t_1$ and $t_2$, and we will study both the imaginary-time and real-time behavior of this quantity. The physics is conveniently discussed in terms of $T/J$ and $J$ as sketched in Figure~\ref{fig:mm_phase_diagram}. Both quantities are dimensionless in units where $\ell$, $m$, and $\hbar$ are set to one (see Appendix~\ref{app:crash_course}). $T\gg J$ is the classical regime and $T \ll J$ is the quantum regime. Moreover, $J \ll 1$ is weak coupling and $J \gg 1$ is strong coupling.  

At zero temperature, $T=0$, we find the imaginary time-dependence $\Delta(\tau) \sim \tau^\alpha$ with $\alpha$ determined by $J$ through
\begin{equation}
2\tan \frac{\pi \alpha}{2}\frac{\pi \alpha}{(\alpha+1)}=J^{-2}
\label{eq:alphaPreview}
\end{equation}
which takes us from the free value $\alpha=1$ at $J^2=0$ to the confined $\alpha=0$ as $J^2$ goes to infinity. In terms of the previously introduced dynamical exponent, we have $z= 2/\alpha$. This identification arises because $\Delta$ scales like two powers of length and length scales like time to the $1/z$ power.

Moreover, we find that the system has reparameterization symmetry like the SYK model, and power-law decay in the velocity correlations. There is also a rescaling symmetry in space, completely separate from the rescaling in time included in the reparameterization mode. This is in contrast with the SYK model, where the fermion modes stretch in a way controlled by the reparameterization mode.

Turning on a non-zero but low temperature, $ T \ll J$, the thermodynamics is controlled by the modified dispersion with non-trivial dynamical exponent, leading to a heat capacity equal to $N/z = N \alpha/2$. For the real time dynamics, we find that the system exhibits properties characteristic of a quantum chaotic system. The velocity auto-correlation now decays exponentially in time with a time-scale set by $\beta = 1/T$. Out-of-time-order correlators also feature an initial period of exponential growth with chaos exponent $\kappa \sim 1/\beta$. As $J$ increases, $\kappa$ gets closer to the chaos bound, $\kappa^{\max} = \frac{2\pi}{\beta}$. Both the relaxation time and the Lyapunov time depend on $J$, so we have in fact an infinite family of ensembles of non-relativistic scale-invariant quantum systems.

At higher temperatures, $T \gg J$, the system becomes effectively classical and the thermodynamics reduces to that of a free particle, corresponding to $z=2$ and $\alpha=1$. The real time dynamics is still chaotic and we compute the diffusivity, the velocity relaxation time, and the largest Lyapunov exponent.

Our detailed analysis of the Magnetic Maze proceeds as follows. In Section \ref{sec:EOM} we use the large-$N$ limit to derive a mean-field action in terms of collective displacement variables. We then derive a saddle-point equation. Throughout Section \ref{sec:EOM} we assume a general correlation function $f$ for the $A$s.

In Section \ref{sec:lowT} we specialize to the situation in equation \eqref{eq:interesting_f}, and work at low temperature. We derive the power-law behavior of the correlation function, as well as equation \eqref{eq:alphaPreview}. We study the low temperature thermodynamics, showing that the low-temperature heat capacity is $\alpha N/2$, which is consistent with the picture of power-law dispersion with exponent $E\propto p^z$ with $z= 2/\alpha$. Finally, we show that at low temperatures and long times, the system has a reparameterization symmetry and a rescaling symmetry.

Section \ref{sec:realTime} covers the real time dynamics of the system. We derive Schwinger-Dyson mean-field equations, and numerically solve them. We show that a particle in the Magnetic Maze exhibits diffusive transport, with the squared displacement after time $t$ growing in proportion to $t$. More precisely, the displacement grows at $2Dt$, where the diffusion constant $D$ scales as $\beta^{\alpha-1}$ at low temperature. In the same temperature regime, the velocity auto-correlation decays exponentially with a time scale set by $\beta$. We also note that the system has a somewhat unusual structure on the Schwinger-Keldysh (SK) contour, as we discuss in detail. 

Section \ref{sec:lyapunov} covers the Lyapunov physics. We setup a ladder calculation of an out-of-time-order correlator (OTOC) and carry out a detailed numerical analysis. We find that the low temperature chaos exponent is of order the bound, $2\pi/\beta$, and comes close to it as $J$ increases.

Section \ref{sec:discussion} discusses directions for further work and open questions. Following it are several technical appendices. 

Appendix~\ref{app:crash_course} provides a crash course on the physics of magnetic fields in high dimensions, analyzes the uniform field problem (which is integrable and exactly solvable at finite $N$), and explains how the dimensionless parameters $T/J$ and $J$ arise.

The other appendices are as follows. Appendix~\ref{app:analyticalJ0} reviews the exact solution for $J=0$. Appendix~\ref{app:other_f} discusses other choices of $f$. Appendix~\ref{suppsec:realtime} presents details of derivation of the real-time equations of motion. Finally, Appendix~\ref{app:numerics} discusses the numerical method used to solve the equations of motion.

\section{Action and Equations of Motion}
\label{sec:EOM}

We are going to perform a mean-field calculation in which the collective variable
\begin{equation}
G(\t_1,\t_2)=\frac 1N \mathcal P \sum x_i(\t_1)x_i(\t_2)
\label{eq:GdefThermo}
\end{equation}
will play a central role. Here, $\mathcal{P}$ denotes an appropriate path- or time-ordering as specified below. Our goal is to find a mean-field expression for the free energy of the system, expressed as a path integral over $G$. 

We start with the partition function
\begin{equation}
    Z=\int \exp\left(\int_0^\beta-\frac m2 v^2-iA\cdot v-\frac \epsilon{2} x^2d\t\right)\mathcal D x.
    \label{eq:xPartition}
\end{equation}
We then introduce the Lagrange multiplier $\Sigma$ enforcing \eqref{eq:GdefThermo}, 
\begin{equation}
    Z=\int \exp\left(\int_0^\beta-\frac m2 v^2-iA\cdot v-\frac \epsilon{2} x^2d\t-\frac 12\int_0^\beta \Sigma(\t_1,\t_2)\{ N G(\t_1,\t_2) -x(\t_1)\cdot x(\t_2)\}d\t_1 d\t_2\right)\mathcal D x \mathcal{D} G \mathcal{D} \Sigma.
    \label{eq:xPartitionSigma}
\end{equation}

Next we perform the average over $A$s to get
\begin{equation}
\begin{split}
    \bar Z&=\int \exp\left(-S\right)\mathcal D x \mathcal{D} G \mathcal{D} \Sigma \\
    S&=\int_0^\beta\frac m2 \dot x^2+\frac \epsilon{2} x^2d\t+\frac 12\int_0^\beta \Sigma(\t_1,\t_2)\left\{NG(\t_1,\t_2) -x(\t_1)\cdot x(\t_2)\right\}+{f\left((x(\t_1)-x(\t_2))^2\right)}v(\t_1)\cdot v(\t_2)d\t_1 d\t_2.
    \label{eq:xPartitionBar}
\end{split}
\end{equation}
Note this leads to an ``annealed'' free energy $\propto \ln \overline{Z}$, whereas the more physical ``quenched'' free energy is $\propto \overline{\ln Z}$. However, as we will see shortly, these free energies always coincide at large $N$ in the Magnetic Maze.

Using the delta function setting $G$ to \eqref{eq:GdefThermo}, we can replace $(x(\tau_1) - x(\tau_2))^2/N = \Delta(\tau_1,\tau_2)$ where
\begin{equation}
    \Delta(\t_1,\t_2)=G(\t_1,\t_1)+G(\t_2,\t_2)-2G(\t_1,\t_2).
\end{equation}
Similarly, we can replace $v(\tau_1) \cdot v(\tau_2)$ with $\partial_{\tau_1} \partial_{\tau_2} G(\tau_1,\tau_2)$. The resulting functional integral is now quadratic in the $x$s, and performing the $x$ integral gives
\begin{equation}
    S/N=\frac 12 \tr \log \left(-m\partial_\t^2+\epsilon-\Sigma\right)+\frac 12 \int \Sigma G+f(\Delta(\t_1,\t_1)){\partial_{\t_1}\partial_{\t_2}G(\t_1,\t_2)}d\t_1d\t_2.
    \label{eq:FullAction}
\end{equation}
This is the sought-for mean-field action which can be used to compute the free energy.

%


%

The equations of motion can now be obtained as functional derivatives of \eqref{eq:FullAction} with respect to $G$ and $\Sigma$. Using the functional derivative
\begin{equation}
    \frac{\delta \Delta(\t_1,\t_2)}{\delta G(\t_1',\t_2')} =  \delta(\t_1 - \t_1')\delta(\t_1 - \t_2') + \delta(\t_2 - \t_1')\delta(\t_2 - \t_2') - 2 \delta(\t_1 - \t_1')\delta(\t_2 - \t_2')
\end{equation}
and the identity $-2\partial_{\t_1}\partial_{\t_2} G = \partial_{\t_1}\partial_{\t_2} \Delta  $, we get the following equations:
\begin{equation}
\begin{split}
(-m\partial_\t^2+\epsilon-\Sigma)*G=\delta(\t_1-\t_2)\\
\Sigma(\t_1,\t_2)=\partial_{\t_1}\partial_{\t_2}(f(\Delta(\t_1,\t_2))+f'(\Delta)\partial_{\t_1}\partial_{\t_2}\Delta-\delta(\t_1-\t_2)\int_0^\beta {f'(\Delta(\t_1,\t'))\partial_{\t_1}\partial_{\t'}\Delta }d\t'.
\end{split}
\label{eq:thermoEOM}
\end{equation}

It is the last term in the equation for $\Sigma$ that is most unfamiliar. It is a consequence of the fact that the interaction term depends not just on $G(\t_1,\t_2)$, but also on $G(\t_1,\t_1)$ and $G(\t_2,\t_2)$. This special dependence of the interaction on $G$ only through its derivatives and through $\Delta$ means that adding an overall constant to $G$ (corresponding to translating the entire path by a random vector) doesn't affect the magnetic part of the action, a consequence of the statistical translational symmetry of our magnetic field. Another consequence is that $\Sigma_{\omega=0}$ is exactly zero.

It is the first equation in \eqref{eq:thermoEOM} that illustrates why we included $\epsilon$. Without it, the zero-frequency mode of $-m\partial_\t^2+\epsilon-\Sigma$ would be zero, and the equation wouldn't be satisfiable. The weak potential from $\epsilon$ saves us from a diverging position by limiting the motion of the particle to a sphere of squared radius $\frac{N}{\epsilon \beta}$. This also cuts out divergences in the partition function and entropy due to the infinite volume of space, instead replacing them with a contribution to the entropy 

\begin{equation}
    \textrm{Entropy from volume of box}=-\frac N2 \log \left(\beta \epsilon\right).
    \label{eq:boxVol}
\end{equation}

It is also useful to write out equations \eqref{eq:thermoEOM} assuming time-translation invariance. In this case, cutting out $G$ entirely, we have
\begin{equation}
\begin{split}
    \Delta_{\omega\neq 0}=\frac{-2}{m\omega^2+\epsilon-\Sigma_{\omega\neq0}}\\
    \Sigma(\t)=-2\partial_\t^2 \Delta f'(\Delta)-(\partial_{\t} \Delta)^2 f''(\Delta)+\delta(\t)\int_0^\beta {f'(\Delta(\t'))\partial_{\t'}^2\Delta(\t') }d\t'.
    \label{eq:thermo_EOM_tt_inv}
\end{split}
\end{equation}

Equations \eqref{eq:thermo_EOM_tt_inv} illustrate why the quenched and annealed free energies are equal in the Magnetic Maze. The quenched free energy can be obtained from a replica trick involving $n$ replicas of the system in the limit $n \to 0$. The mean-field variables would be generalized to $G_{ab}$, $\Delta_{ab}$, and $\Sigma_{ab}$, where $a,b=1,\cdots,n$ are replica indices. Replica symmetry breaking requires correlations between the replicas, but the inter-replica correlations are typically time-independent owing to the separate time-translation symmetry in each replica. The generalization of \eqref{eq:thermo_EOM_tt_inv} to the multi-replica case would relate inter-replica terms in the self-energy, $\Sigma_{a \neq b}$, to time derivatives of the inter-replica terms in the displacement, $\Delta_{a \neq b}$. Since the inter-replica correlators are time-independent, we can conclude from the generalization of the second equation of \eqref{eq:thermo_EOM_tt_inv} that $\Sigma_{a \neq b} = 0$. Self-consistency from the multi-replica generalization of the first equation of \eqref{eq:thermo_EOM_tt_inv} then implies that $\Delta_{a \neq b}=0$. Hence, multi-replica saddle points are always replica diagonal in the Magnetic Maze.

\section{Low-Temperature Limit with $f\sim \frac{J^2}{\Delta}$}
\label{sec:lowT}

In this section, we study the thermodynamics of the Magnetic Maze with $f(\Delta) \sim\frac{J^2}{\Delta}$ and reveal its fascinating low-temperature physics. In the IR limit, where $\Delta$ is large and $\partial_\t$ is small, we can simplify our action. Equation \eqref{eq:FullAction} becomes
\begin{equation}
    S/N=\frac 12 \tr \log \left(\Sigma\right)-\frac 14 \int \left( \Sigma \Delta +J^2\frac{\partial_{\t_1}\partial_{\t_2}\Delta(\t_1,\t_2)}{\Delta(\t_1,\t_1))} \right) d\t_1d\t_2.
    \label{eq:IRAction}
\end{equation}
This IR action behaves in a simple way under two important classes of transformations: time reparameterizations and scaling transformations.

For reparameterizations of time, $\t \to \sigma(\t)$, the transformations are
\begin{align}
     G(\t_1,\t_2) &\to G(\sigma(\t_1),\sigma(\t_2)) \\
      \Delta(\t_1,\t_2) &\to \Delta(\sigma(\t_1),\sigma(\t_2)) \\
      \Sigma(\t_1,\t_2) & \to \Sigma(\sigma(\t_1),\sigma(\t_2)) |\sigma'(\t_1) \sigma'(\t_2)|.
\end{align}
It is straightforward to verify that this transformation preserves the last two terms in \eqref{eq:IRAction}. The $\det \ln \Sigma$ term is not manifestly symmetric, but by writing the transformation as
\begin{equation}
    \Sigma(\t_1,\t_2) \to \int d\t_1' d\t_2' M(\t_1,\t_1') \Sigma(\t_1',\t_2') M(\t_2',\t_2)
\end{equation}
for a ``matrix'' $M(\t_1,\t_1') = \delta(\t_1' - \sigma(\t_1)) | \sigma'(\t_1)|$, we can argue that the determinant is invariant by arguing that the determinant of $M$ is unity.\footnote{For example, if $\sigma$ is a time shift, $\sigma(\t)=\t+a$, then $M = \delta(\t_1' - \t_1-a)$. This is effectively a permutation, which has determinant one. For a general infinitestisimal transformation, $\sigma = \t + \epsilon(\t)$, one can directly show that the determinant is one by using the single-valuedness of $\epsilon(\t)$.} 

For scalings, the time is unchanged but we send
\begin{align}
    G &\to \lambda G \\
    \Delta &\to \lambda \Delta \\
    \Sigma &\to \lambda^{-1} \Sigma.
\end{align}
We can also extend this transformation to the underlying $x(\t)$ variables as $x(\t) \to \sqrt{\lambda} x(\t)$. The last two terms in \eqref{eq:IRAction} are again manifestly invariant, but the first term transforms by a field-independent shift that depends on $\lambda$. As a consequence, the equations of motion in the IR limit are invariant under the scaling transformation but the action is not. This situation is slightly unusual but it can be understood as follows. Going back to \eqref{eq:xPartitionBar} before the $x$ variables were integrated out, we see that the action is fully invariant under the scaling transformation provided we drop the corresponding terms (the $x$ kinetic term, the confining potential, and the short-distance part of $f$), but the $\mathcal{D}x$ measure is not invariant (while the combined $\mathcal{D}G \mathcal{D}\Sigma$ measure is). The scaling symmetry is also explicitly broken by the aforementioned terms we neglected in the IR limit.

Both of these transformations are approximate symmetries of the action in the IR limit, \eqref{eq:IRAction}, with the caveat about scalings discussed just above. However, we will see in the next subsection that the ground state spontaneously breaks these symmetries to a special subgroup.

\begin{figure}
    \centering
    \includegraphics[width=0.75\linewidth]{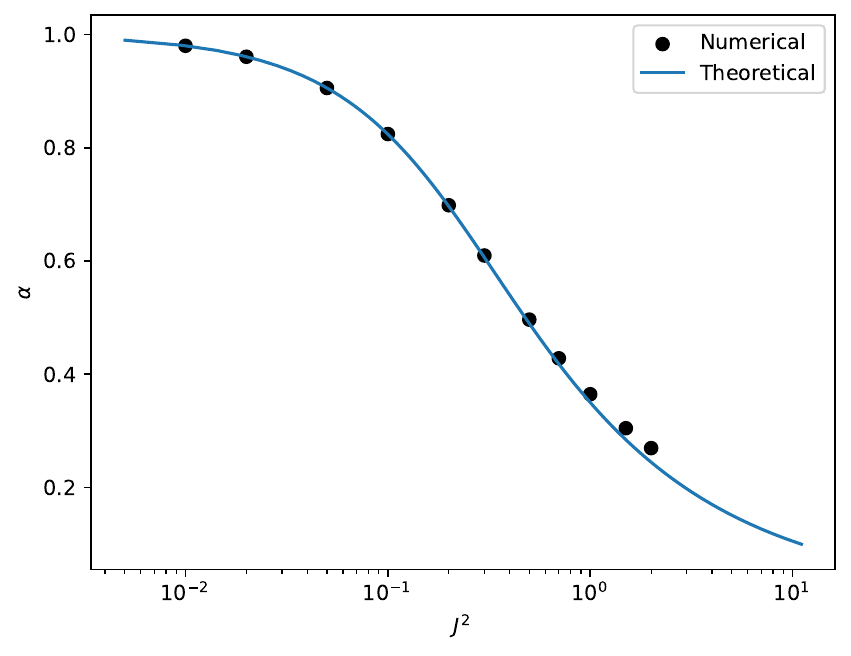}
    \caption{(Blue) The theoretical value of $\alpha$ extracted from equation \eqref{eq:JAlpha}. (Black) The value of $\alpha$ extracted from comparing the maximum value to $\Delta$ when $\beta=5000$ vs $10000$. We see the fit begin to fail at large values of $J^2$ when the assumption of large $\Delta$ isn't satisfied even when $\beta$ is in the thousands.}
    \label{fig:alphaScatter}
\end{figure}

\subsection{Power-Law Behavior of Ground State Correlations}

We now show that a power-law ansatz solves the equations of motion arising from \eqref{eq:IRAction} in the zero temperature limit. We consider a translation invariant ansatz in which $\Delta(\t_1,\t_2) = \Delta(\t_1-\t_2)$ and similarly for $\Sigma$.

Suppose the IR behavior of $\Delta(\tau)$ is 
\begin{equation}
 \Delta(\tau)= k|\tau|^\alpha.   
\end{equation}
Then the IR value of $\Sigma$ is given by 
\begin{equation}\Sigma(\tau)=-\frac{J^2 2\alpha}{k} |\tau|^{-\alpha-2}.\end{equation} 

This gives 
\begin{equation}
\begin{split}
\Delta_\omega=-2\sin \frac{\pi \alpha}{2}\Gamma(\alpha+1)k |\omega|^{-\alpha-1}\\
\Sigma_{\omega}=2\sin \frac{\pi \alpha}{2}\Gamma(-\alpha-1)\frac{J^2 2\alpha}{k} |\omega|^{\alpha+1},
\end{split}
\end{equation}
as well as
\begin{equation}
\Delta_\omega \Sigma_{\omega}=-8\sin^2 \frac{\pi \alpha}{2}\Gamma(-\alpha-1)\Gamma(+\alpha+1)\alpha J^2=-2.
\end{equation}
We can use the reflection formula,
\begin{equation}
    \Gamma(z)\Gamma(-z)=\frac{-\pi}{z\sin(\pi z)},
\end{equation}
to get
\begin{equation}
2\tan \frac{\pi \alpha}{2}\frac{\pi \alpha}{(\alpha+1)}=J^{-2}.
\label{eq:JAlpha}
\end{equation}

It turns out that for any $J$ we pick, there is a unique $0<\alpha<1$ which solves equation \eqref{eq:JAlpha}. We see in figure \ref{fig:alphaScatter} that equation \eqref{eq:JAlpha} aligns perfectly with numerics for reasonably small $J^2$ at a certain large fixed $\beta$. For larger $J^2$, one would need correspondingly larger $\beta$ to see this agreement.

From the explicit form $\Delta \sim \tau^\alpha$, we learn that time reparameterizations are broken down to translations. Moreover, the scaling symmetry is also broken. However, the combination of a time reparameterization that rescales time and a scaling transformation leaves the form of $\Delta$ invariant. In this sense, the symmetries discussed in the previous subsection are spontaneously broken in the ground state to a subgroup of translations and combined rescalings. Of course, these symmetries are also explicitly broken by the non-IR terms in the full action.

\subsection{Imaginary Time Numerics}\label{ssub:Imag_time_numeircs}

\begin{figure}
    \centering
    \includegraphics[width=0.95\linewidth]{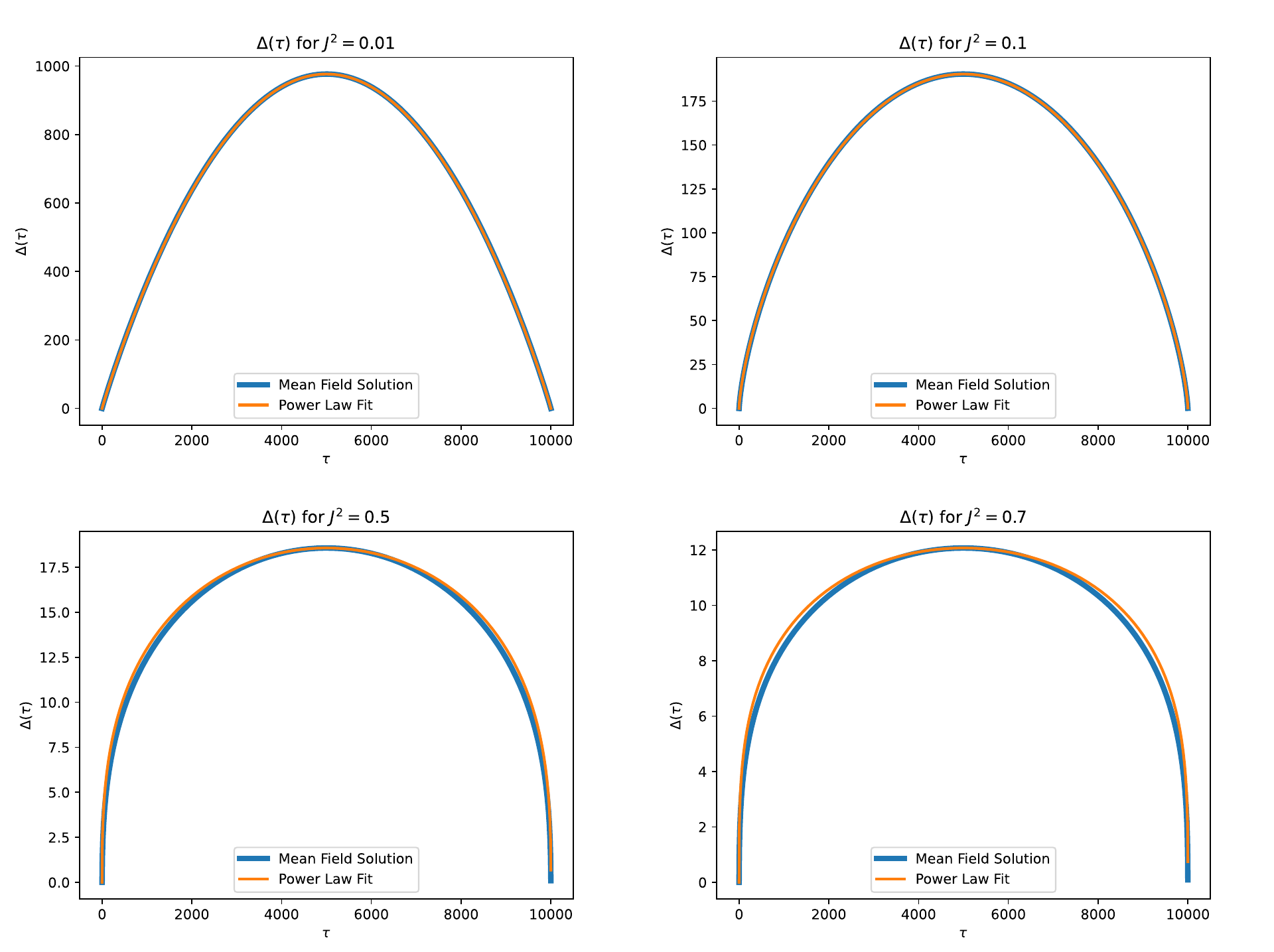}
    \caption{Numerical solution to the mean field equations in blue, compared with equation \ref{eq:DeltaAnsatz} chosen to match the value and curvature of $\Delta$ at $\t=\beta/2$. We see strong agreement between the two curves as long as we are in the region where $\Delta \gg 1$. At larger $J^2$, however, this particular form seems to break down, although $\Delta(\t)$ is certainly proportional to $\t^\alpha$ for $\t\ll \beta$ and $(\beta-\t)^\alpha$ for $\beta-\t\ll \beta$. See the discussion around equation~\eqref{eq:Delta_even_more_general}.}
    \label{fig:coldNumerics}
\end{figure}

We saw in the previous subsection that at zero temperature, $\Delta(\t)$ is proportional to $\t^\alpha$ with a specific $\alpha$ given by the solution to the transcendental equation \eqref{eq:JAlpha}. By analogy with the SYK model, one might expect the finite-temperature analog to be $\Delta(\t)\propto \left(\sin \frac{\pi \t}{\beta}\right)^\alpha$. In fact, we see from the graphs in Figure \ref{fig:coldNumerics} that for small $J^2$ the solution looks like
\begin{equation}
    \Delta(\t)\propto\left(\t(\beta-\t)\right)^\alpha.
    \label{eq:DeltaAnsatz}
\end{equation}
When $J^2=0$ and $\alpha=1$, this ansatz becomes exact (see appendix \ref{app:analyticalJ0}). At large $J^2$, the magnetic length becomes much smaller than $\ell$. This maps onto the constant field limit discussed in Appendix \ref{app:crash_course}, where the large-$\beta$ solution is found to be proportional to $\log \beta \sin \frac{\pi \t}{\beta}$ for large $\beta$. This is not the small-$\alpha$ limit of equation \eqref{eq:Delta_ansatz}, suggesting that the function takes the more general form
\begin{equation}
    \Delta(\t)=\left(\beta F_\alpha\left(\frac{\t}{\beta}\right)\right)^\alpha \label{eq:Delta_even_more_general}
\end{equation}
where $F_\alpha$ is some $\alpha$ dependent function satisfying $F_\alpha(x)=F_\alpha(1-x)$ (by the KMS condition \cite{PhysRev.115.1342,doi:10.1143/JPSJ.12.570}) and $F_\alpha(x)\sim~x$ for small $x$. For $\alpha\sim~1$ $F_\alpha(x)$ seems to take the form $x(1-x)$, while for $\alpha\sim 0$ we have $F_\alpha(x)=\sin \pi x.$

\subsection{The Free Energy at Low Temperature}
To cap off our discussion of thermodynamics, we will calculate the free energy of our system at low temperatures.

As already discussed in equation \eqref{eq:boxVol}, there is a contribution to the entropy of $-\frac N2 \log \left(\beta \epsilon\right)$, and a corresponding contribution to the free energy of $\frac N2 \beta^{-1}\log \left(\beta \epsilon\right)$, essentially the log-volume of the "box" the particle is in.

The most physically meaningful quantities are the ones that don't depend on the box volume: energy and heat capacity. Even more meaningful is kinetic energy. The potential energy is an artifact of our mathematically convenient choice to confine our system with a quadratic potential; other shapes such as quartic or infinite well would give different potential energies, but the kinetic energy would be invariant so long as the box is large.

One can see by examination that Hamiltonian \eqref{eq:bigHam} is always positive. Furthermore, at high temperatures the $A$s become negligible compared to the $p$s, and our particle becomes essentially classical. A classical particle in a magnetic field will have kinetic energy $N\beta^{-1}/2$ by the equipartition theorem.
    \label{eq:FullActionAgain}

\begin{figure}
    \centering
    \includegraphics[width=0.9\linewidth]{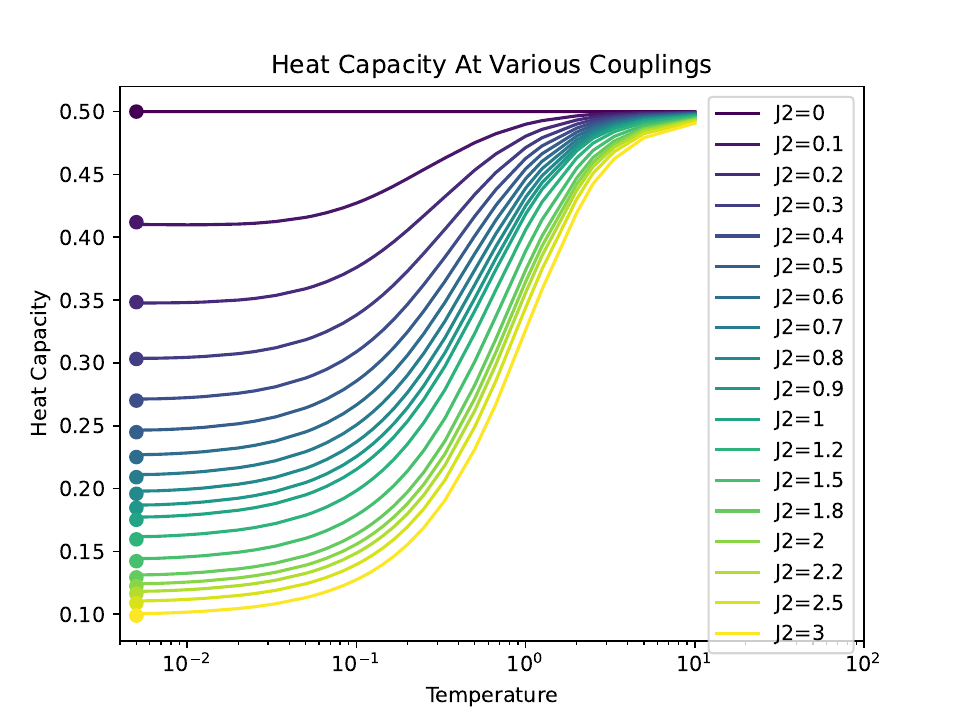}
    \caption{As the temperature goes from low to high, the heat capacity smoothly interpolates from $\alpha(J^2)/2$ to $1/2$ for a wide range of $J^2$s. The colored lines show the heat capacity $\frac{\partial \textrm{Kinetic Energy}}{\partial T}$ for different values of $J^2$, while the corresponding dots show the predicted low-temperature heat capacities.}
    \label{fig:heatCapacities}
\end{figure}

At low temperatures, there is an equally natural interpretation of the thermodynamics. The imaginary-time distance function $\Delta(\t)\sim\t^\alpha$ corresponds to a particle with dispersion $E\sim|p|^{2/\alpha}$. Working out the thermodynamics of such a particle, we see that the entropy goes as $S=\alpha \frac N2 \log \beta+\textrm{const}$, and the heat capacity is precisely $\alpha \frac N 2$. We see in Figure \ref{fig:heatCapacities} that the intensive heat capacity is indeed $\alpha/2$ at low temperatures.

\section{Real-Time Dynamics}
\label{sec:realTime}

In this section, we study the real-time dynamics of the magnetic maze. We firstly set up the Schwinger-Keldysh formalism and derive the saddle point equations. Secondly, we solve these equations for specific vector potential distribution Eq.~\eqref{eq:interesting_f}. The system shows a universal long-time limit in which the particle motion is diffusive.  

        \subsection{Schwinger-Keldysh Formulism}
        \subsubsection{Derivation of the Saddle Point Equations}

        To study the two-point correlation function in real-time, it is convenient to represent the time-evolved density matrix using the Keldysh contour [Kamenev, Altland]
	\begin{equation}
		Z = \Tr(U_{-\infty,\infty} U_{\infty,-\infty}\rho_0)=\int \mathcal{D} x  \exp\left(\iu \sum_j \sum_{a} (-1)^{a}\int_{-\infty}^{\infty} \diff t \left( \frac m2 \dot{x}_{a,j}^2 +  A(x_{a,j})\cdot \dot{x}_{a,j} -\frac{\epsilon}{2} x_{a,j}^2 \right) \right),
		\label{eq:xPartition}
	\end{equation}
        where $a = +, -$ indicates the forward and backward contours, respectively, with the phase $(-1)^{a}$ representing the direction of unitary evolution. By convention, we assume $a = 0, 1$ corresponds to the contour labels $+, -$. Here, $\rho_0 = e^{-\beta H}$ is the thermal density matrix. For the real-time dynamics, the corresponding Green's function is defined as follows:
	\begin{equation}
		\iu G_{ab}(t_1,t_2)=\frac{1}{N} \mathcal{P} \sum_{i} x_{a,i}(t_1)x_{b,i}(t_2).
		\label{eq:Gdef}
	\end{equation}
	
	We introduce the Lagrange multiplier $\Sigma_{ab}$ to enforce Eq.~\eqref{eq:Gdef}
\begin{equation}\label{eq:xPartitionSigma}
		\begin{split}
			Z&=\int \mathcal{D} x  \exp\Bigg(\iu \sum_j \sum_{a} (-1)^{a}\int_{-\infty}^{\infty} \diff t \left( \frac m2 \dot{x}_{a,j}^2 + A(x_{a,j})\cdot \dot{x}_{a,j} -\frac{\epsilon}{2} x_{a,j}^2 \right) \\
			& - \sum_{ab} (-1)^{a+b}  \frac{1}{2} \int \diff t_1 \diff t_2 \left[ N\Sigma_{ab}(t_1,t_2)G_{ab}(t_1,t_2) + \iu \sum_j \Sigma_{ab}(t_1,t_2)x_{a,j}(t_1)\cdot x_{b,j}(t_2) \right] \Bigg) \\
		\end{split}
	\end{equation}
	Similar to the imaginary-time derivation, we average over the $A$ fields to obtain
	\begin{equation}
		\begin{split}
			\bar Z= &\int \exp\left(\iu S\right)\mathcal{D} x\\
			\iu S/N =&  \iu \sum_{a} (-1)^{a}\int \diff t_1 \diff t_2  x_{a}(t_1) \left( -\frac m2 \partial_{t_1}\partial_{t_2}\delta(t_1-t_2)\delta_{ab}  -\frac{\epsilon}{2} \delta(t_1-t_2)\delta_{ab} - \sum_{b}(-1)^{b}\frac{\Sigma_{ab}(t_1,t_2)}{2} \right) x_{b}(t_2) \\ 
			& - \sum_{ab} (-1)^{a+b}  \int \diff t_1 \diff t_2  \left( \frac{1}{2} \Sigma_{ab}(t_1,t_2)G_{ab}(t_1,t_2) + \frac{1}{2}  f\left((x_a(t_1)-x_b(t_2))^2/N\right) \dot{x}_a(t_1)\cdot \dot{x}_b(t_2)/N \right) \\
			\label{eq:xPartitionBar_Real}
		\end{split}
	\end{equation}
	
        We can perform the quadratic integral over the $x_a$, which leads to the effective action.
	\begin{equation}
		\begin{split}
			\iu S/N=&  - \frac 12 \log \det 
			\left( (-1)^{a}(- m \partial_{t_1}\partial_{t_2}-\epsilon)\delta(t_1-t_2)\delta_{ab} - (-1)^{a+b}\Sigma_{ab} \right)  \\
		&- \sum_{ab} (-1)^{a+b}  \int \diff t_1 \diff t_2  \left( \frac{1}{2} \Sigma_{ab}(t_1,t_2)G_{ab}(t_1,t_2) + \frac{1}{2} f\left(\Delta_{ab}(t_1,t_2) \right) \partial_{t_1}\partial_{t_2} \iu G_{ab}(t_1, t_2) \right) \\
		\end{split}
		\label{eq:FullActionReal}
	\end{equation}
	where we defined 
	\begin{equation}\label{eq:Deltarealtime_def}
		\Delta_{ab}(t_1,t_2)=\iu \left(G_{aa}(t_1,t_1)+G_{bb}(t_2,t_2)-G_{ab}(t_1,t_2) - G_{ba}(t_2,t_1)\right)
	\end{equation}
	to be the squared distance between the positions at $t_1$ and $t_2$. In Eq.~\eqref{eq:FullAction}, all fields are defined on the original Keldysh contour. It is convenient to obtain the relation between the self-energy and Green's function by using the first set of saddle point equations, $\frac{\partial S}{\partial G_{ab}} = 0$, which reads
    \begin{equation}\label{eq:selfE_ab}
        \begin{split}
			&\Sigma_{ab}(t_1,t_2) =-\iu \partial_{t_1}\partial_{t_2}f(\Delta_{ab}(t_1,t_2)) - \iu f'(\Delta_{ab}(t_1,t_2))\partial_{t_1}\partial_{t_2}\Delta_{ab}(t_1,t_2) + \delta(t_{12})\delta_{ab} \int \diff{t_3} \Bigg(   \\
    &\ \  \iu f'(\Delta_{ab}(t_1,t_3))\partial_{t_1}\partial_{t_3}\Delta_{ab}(t_1,t_3) -\frac{1}{2}\iu f'(\Delta_{+-}(t_1,t_3))\partial_{t_1}\partial_{t_3}\Delta_{+-}(t_1,t_3) - \frac{1}{2}\iu f'(\Delta_{-+}(t_1,t_3))\partial_{t_1}\partial_{t_3}\Delta_{-+}(t_1,t_3) \Bigg)\\
        \end{split}
    \end{equation}
    where $G_{ab}\equiv \begin{pmatrix}
        G_T & G_< \\
        G_> & G_{\tilde{T}} \\
    \end{pmatrix}_{ab}$ and the same definition applies for $\Sigma_{ab}$. Here, $T$ and $\tilde{T}$ denote time-ordering and anti-time-ordering, respectively, while $<,>$ label the lesser and greater Green's functions, as the two time variables are on different contours.
    
    The Schwinger-Dyson equations are obtained by computing $\frac{\partial S}{\partial \Sigma_{ab}}$, with $a,b$ either in the basis $+,-$ or $cl,q$. In the basis $x_{cl}, x_q$ after the Keldysh rotation, the Schwinger-Dyson equations read
    \begin{equation}\label{eq:SD_full}
        \begin{pmatrix}
				0 & -\delta(t_{12}) \left(m\partial_{t_2}^2+\epsilon\right)-\Sigma_A \\
				-\delta(t_{12}) \left(m\partial_{t_2}^2+\epsilon\right)-\Sigma_R & -\Sigma_K \\
        \end{pmatrix} \circ \begin{pmatrix}
			G_K & G_R \\
			G_A & 0  \\
		\end{pmatrix} = \mathbb{I}
    \end{equation}
	Here, the $\circ$ symbol denotes convolution: $A(t_1, t_2) \circ B(t_2, t_3) \equiv \int \diff t_2 A(t_1, t_2) B(t_2, t_3)$. It is also useful to express Eq.~\eqref{eq:SD_full} under the assumption of time-translation invariance and perform a Fourier transformation. The detailed derivation is provided in Appendix \ref{suppsec:realtime}, and the self-energy is summarized in Table \ref{tab:realtime_selfE}. We obtain the retarded component of the Schwinger-Dyson equation as follows: 
\begin{equation}\label{eq:Schwinger_Dyson}
		\begin{split}
			&\left(m(\omega+ \iu \eta)^2 - \epsilon-\left(\bar{\Sigma}_R(\omega)-\bar{\Sigma}_R(\omega=0)\right)\right)  G_R(\omega)=1\\
			&\bar{\Sigma}_{R}(t)= \Theta(t)\left(\Sigma_{>}(t)-\Sigma_{<}(t)\right)  \\        
		\end{split}
	\end{equation}
    The central result from the derivation is that $\Sigma_R(\omega) = \bar{\Sigma}_R(\omega) - \bar{\Sigma}_R(\omega=0)$. This exactly satisfies the condition $\Sigma_R(\omega=0)=0$, as expected from the imaginary time calculation. The retarded Green's function generally requires a small cutoff $\eta$ to retain the retarded causal structure. In principle, $\eta$ should be taken as an infinitesimal number, $\eta = 0^+$, but a finite $\eta$ is necessary in numerics to help with the convergence of the system, thus introducing artificial dissipation. In this case, we take $\epsilon$ to be zero while keeping the retarded cutoff $\eta$. Consequently, we focus only on the real-time Green's function where the time variable $t \ll 1/\eta$ to eliminate unphysical artifacts. This sets a lower bound for the low-temperature region, requiring the inverse temperature $\beta \ll 1/\eta$.
	
	To self-consistently solve the equation, we also require the fluctuation-dissipation theorem to determine the relation between $G_R$ and $G_{\gtrless}$,
    \begin{equation}\label{eq:GF_FDT}
		\begin{split}
			\rho_G(\omega) &= -\frac{1}{2\pi \iu} (G_R(\omega) - G_R(\omega)^{\dagger}) \\
			G_<(\omega) &= -2\pi \iu n_B(\omega) \rho(\omega) \\
			G_>(\omega) &= +2\pi \iu n_B(-\omega) \rho(\omega), \\
		\end{split}
	\end{equation}
    where $n_B(\omega) = 1/\left(e^{\beta \omega} - 1\right)$ is the bosonic distribution function and $\beta$ is the inverse temperature.

Note that Green's function $G$ is not well-defined without a trapping potential $\epsilon$. An example of this is the analytical solution for free particles at $J=0$, as shown in Appendix \ref{app:analyticalJ0}. For the real-time Green's functions $G_{\gtrless}(t)$ and $G_K(t)$, as well as the imaginary-time-ordered Green's function $G(\tau)$, all of them diverge in the limit $\epsilon \to 0$. However, $\Delta_{\gtrless}(t)$, $\Delta_K(t)$, or $\Delta(\tau)$, which represent the squared displacement, are well-defined. Therefore, it is convenient to relate everything to the correlation function $\Delta$.
From the definition of $\Delta_{ab}$ in Eq.~\eqref{eq:Deltarealtime_def}, we can assume time-translation invariance and transform to the frequency domain. Using the symmetry $\Delta_{ab}(t) = \Delta_{ba}(-t)$, we obtain:
\begin{equation}
    \Delta_{ab}(\omega) =  (\text{const.})\,\delta(\omega) -2 \iu G_{ab}(\omega).
\end{equation}
Hence, up to a constant shift at zero frequency, $\Delta_{ab}$ and $G_{ab}(\omega)$ are related by $-2\iu$. The constant shift can be ignored since we always enforce the condition $\Delta_{ab}(t=0)=0$ by taking $\Delta_{ab}(\omega = 0) = - \sum_{\omega \neq 0} \Delta_{ab}(\omega)$ in the numerics.
Consequently, we can define $\rho_{\Delta} = - \frac{1}{2\pi \iu} \left(\Delta_R(\omega) - \Delta_A(\omega)\right)$ for $\omega \neq 0$, and the corresponding relation between the retarded and advanced components is given by
\begin{equation}
    \Delta_R^{*}(\omega) \equiv -\Delta_A(\omega),\ \Delta_R^{*}(t) \equiv -\Delta_A(-t).
\end{equation}
Here, the $*$ symbol denotes taking the complex conjugate, and the interchange of time variables has been carried out explicitly. The spectral function $\rho_{\Delta}$ also satisfies the corresponding fluctuation-dissipation theorem for the correlator $\Delta$. 

In summary, we have the full set of equations for $\Delta$ in Eq.~\eqref{eq:Delta_FDT1}-\eqref{eq:SD_R_Delta}. These can, in principle, be self-consistently solved to yield all real-time Green's functions by providing an initial guess for $\Delta_R(\omega)$.
\boxalign[0.98\textwidth]{
\begin{align}
    \rho_{\Delta}(\omega) &= -\frac{1}{2\pi \iu} (\Delta_R(\omega) - \Delta_A(\omega)),\ \ \omega\neq 0 \label{eq:Delta_FDT1}\\
    \Delta_<(\omega) &= -2\pi \iu n_B(\omega) \rho_{\Delta}(\omega),\ \ \omega\neq 0 \label{eq:Delta_FDT2}\\
    \Delta_>(\omega) &= +2\pi \iu n_B(-\omega) \rho_{\Delta}(\omega),\ \ \omega\neq 0 \label{eq:Delta_FDT3}\\
    \Sigma_{\gtrless}(t) &= \iu \partial_{t}^2 f(\Delta_{\gtrless}(t)) + \iu f'(\Delta_{\gtrless}(t))\partial_{t}^2 \Delta_{\gtrless}(t) \label{eq:DefSigmapm}\\
    \bar{\Sigma}_R(t) &= \Theta(t) \left(\Sigma_>(t)- \Sigma_<(t)\right) \label{eq:DefSigmaR}\\ 
     \Big(m(\omega+& \iu  \eta)^2 - \epsilon-\left(\bar{\Sigma}_R(\omega)-\bar{\Sigma}_R(\omega=0) \right) \Big)  \Delta^R(\omega)=-2\iu ,\ \ \omega\neq 0  \label{eq:SD_R_Delta}
\end{align}
}

\subsubsection{Numerical Methods to Obtain the Solution}\label{ssub:numerical_method}
Here we explain how the real-time equations can be solved numerically. Since the value of $\Delta_{\gtrless}(\omega = 0)$ is undetermined in the self-consistent process, using the simple mixing method $\Delta_R^{(n)}=(1-\zeta) \Delta_R^{(n-1)}+\zeta\Delta_{R,\text{new}}^{(n-1)}$ is numerically unstable. Here, $n$ is the iteration step and $\Delta_{R,\text{new}}^{(n-1)}$ is calculated through the Schwinger-Dyson equation Eq.~\eqref{eq:SD_R_Delta} based on the solution $\Delta_R^{(n-1)}$. Instead, we choose to use a gradient descent protocol to achieve self-consistency.

If we only consider the retarded component, we will find that the gradient being zero is the same as the Schwinger-Dyson equation Eq.~\eqref{eq:SD_R_Delta}. Leaving the details to the supplementary material, we choose the update program to be:
\begin{equation}
    \begin{split}
        \Delta(\omega)_R^{(n))} = \Delta(\omega)_R^{(n-1)} + \zeta \frac{\iu}{4}  \left( (-2\iu)(\Delta_{R}^{(n-1))})^{-1}(\omega) - (G_{0,R}^{-1}(\omega) - \Sigma_R^{(n-1)}(\omega))  \right) \left(\Delta_R^{(n-1)}(\omega) \right)^2
    \end{split}
\end{equation}
Here, the second term $ \frac{\iu}{4}  \left( (-2\iu)\Delta_{R}^{-1}(\omega) - (G_{0,R}^{-1}(\omega) - \Sigma_R(\omega))  \right) = \frac{\partial S[\Delta]}{\partial \Delta_R}$ is the gradient of the action with respect to the variable $\Delta_R$. We also introduce an extra $\Delta_R(\omega)$ as a numerical trick, which stabilizes the iteration process. This is because $\Delta_R^{-1}$, $\Sigma_R$, and $G_{0,R}^{-1}$ all approach zero as $\omega \to 0$, and $\Delta_R^{2}$ enhances the numerical difference around $\omega \approx 0$. We have checked that convergence is reached when $||\Delta_{R}^{(n)}(\omega)-\Delta_{R}^{(n-1)}(\omega)||_2 < 10^{-6} ||\Delta_{R}^{(n)}(\omega)||_2$.

After obtaining the numerical solution, we can check the real-time dynamics by directly comparing them with the imaginary time dynamics as a benchmark. The correlator can be related to the spectral function by the spectral representation
\begin{equation}
    \Delta_R(\omega) = \int \diff{\omega'} \frac{\rho_{\Delta}(\omega')}{\omega' - (\omega + \iu \eta)}.
\end{equation}
In real-time dynamics, we can directly sample the imaginary time Green's function through the analytical continuation
\begin{equation}
    \Delta(\iu \omega_n) = \int \diff{\omega'} \frac{\rho_{\Delta}(\omega')}{\omega' - \iu \omega_n},
\end{equation}
with the Fourier transformation
\begin{equation}
    \Delta(\tau) = \sum_{\omega_n=2\pi n/\beta} \Delta(\iu \omega_n) e^{\iu \omega_n \tau}.
\end{equation}


\subsection{Long-Time Limit of The Dynamics}

\begin{figure}[htb]
\centering
\includegraphics[scale=0.62]{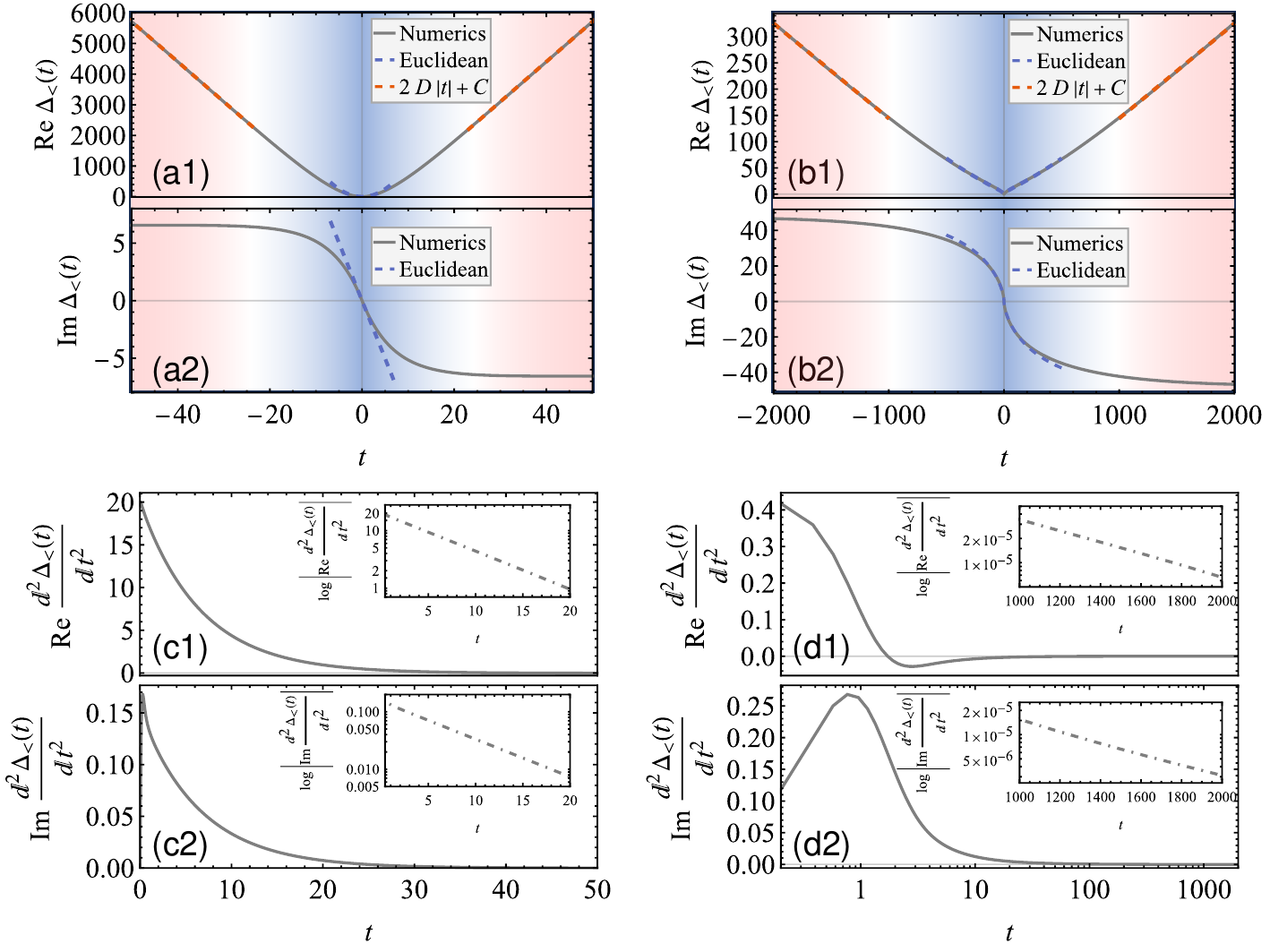}
\caption{ The numerical results of the real-time correlator $\Delta_<(t)$ are shown for both high and low temperatures. (a1, a2, c1, c2) The high-temperature solution is obtained at $\beta=0.1$ and $J^2=0.3$. (b1, b2, d1, d2) The low-temperature solution is obtained at $\beta=500$ and $J^2=0.3$. In the numerical calculations, we set $l=1$, $\eta=5\times 10^{-4}$, and the numerical grid number $N_t=2^{20}$. The total time-domain cutoff is $t_{\text{cut}}=30/\eta$ for high temperature or $t_{\text{cut}}=50/\eta$ for low temperature. 
In Fig.~(a1,a2) and (b1,b2), we demonstrate the real-time solution alongside the Euclidean ansatz form $c_{\alpha} \left(\frac{t(t-\iu \beta)}{\beta}\right)^{\alpha}$, which mostly agree in the region $t\leq \beta$, as indicated by the blue dashed line and blue background color. After this region, determined by $\alpha$, the correlator eventually relaxes to linear growth behavior, or equivalently, diffusion behavior with $\Delta_<(t)\approx 2Dt + C$, as indicated by the red dashed line and red background color. 
To further elaborate on the diffusion behavior, in Fig.~(c1, c2) and (d1, d2), we examine the second derivative $\frac{\diff^2 \Delta_<(t)}{\diff t^2}$, which indicates that the second derivative decays at late times. Remarkably, the insets in Fig.~(c1, c2) and (d1, d2) show that the relaxation is exponential, and we can extract the relaxation time $\rel$ by linear fitting in the logarithmic plot.}
\label{fig:highandlowT_numerics}
\end{figure}

Here we describe the structure of the real-time dynamics in equilibrium at inverse temperature $\beta$. We demonstrate two typical numerical results in Fig.~\ref{fig:highandlowT_numerics}, with (a1,a2,c1,c2) in the high temperature regime and (b1,b2,d1,d2) in the low temperature regime. From this data, it is apparent that the equilibrium real-time dynamics exhibits a rich structure. Here we show that, for all parameters we studied, the dynamics eventually enters into a late-time ``hydrodynamic'' regime in which the large scale motion is diffusive, $\Delta_<(t) \to  2D t$, with $D$ the diffusivity. Moreover, the system relaxes to this diffusive behavior exponentially with a time constant $t_{\text{rel}}$. The time-scale for the onset of the late-time hydrodynamic regime is also of order $t_{\text{rel}}$. 

Before proceeding, we note that the dynamics at early time is quite complex. For example, the Euclidean scaling form, Eq.~\eqref{eq:DeltaAnsatz}, has an imprint on the early time dynamics at very low temperature (see (b1,b2) of Fig.~\ref{fig:highandlowT_numerics}). Nevertheless, we focus on the long-time dynamics as it is the most universal and it is closely related to existence of a non-vanishing chaos exponent (discussed in Sec.~\ref{sec:lyapunov}).

The main goal of the rest of this subsection is to explain our results for $D$ and $t_{\text{rel}}$ at very high and very low temperature. At low temperature, we find $t_{\text{rel}} \sim \beta$ and $D \sim \beta^{2/z-1}$. These low temperature forms are determined by the emergent scaling symmetry and the associated dynamical exponent $z$, but we have not been able to give an analytic derivation of the coefficients. At high temperature, we find $t_{\text{rel}} \sim \beta^{-1/2} J^{-2}$ and $D \sim \beta^{-3/2} J^{-2}$. These we will derive analytically below.

The values for $D$ and $t_{\text{rel}}$ just quoted, and the remainder of the discussion in this subsection, all consider $f \sim J^2/\Delta$. For other choices of $f$, we expect the system will still exhibit exponential relaxation to diffusive dynamics at long time (apart from the special case of uniform magnetic field or zero coupling). It is also interesting to consider the possibility that the dynamics could localize in some fashion, but we have not seen evidence for this in our numerics.

\subsubsection{Theory of Low-Temperature Parameters}

\begin{figure}[t]
\centering
\includegraphics[scale=0.5]{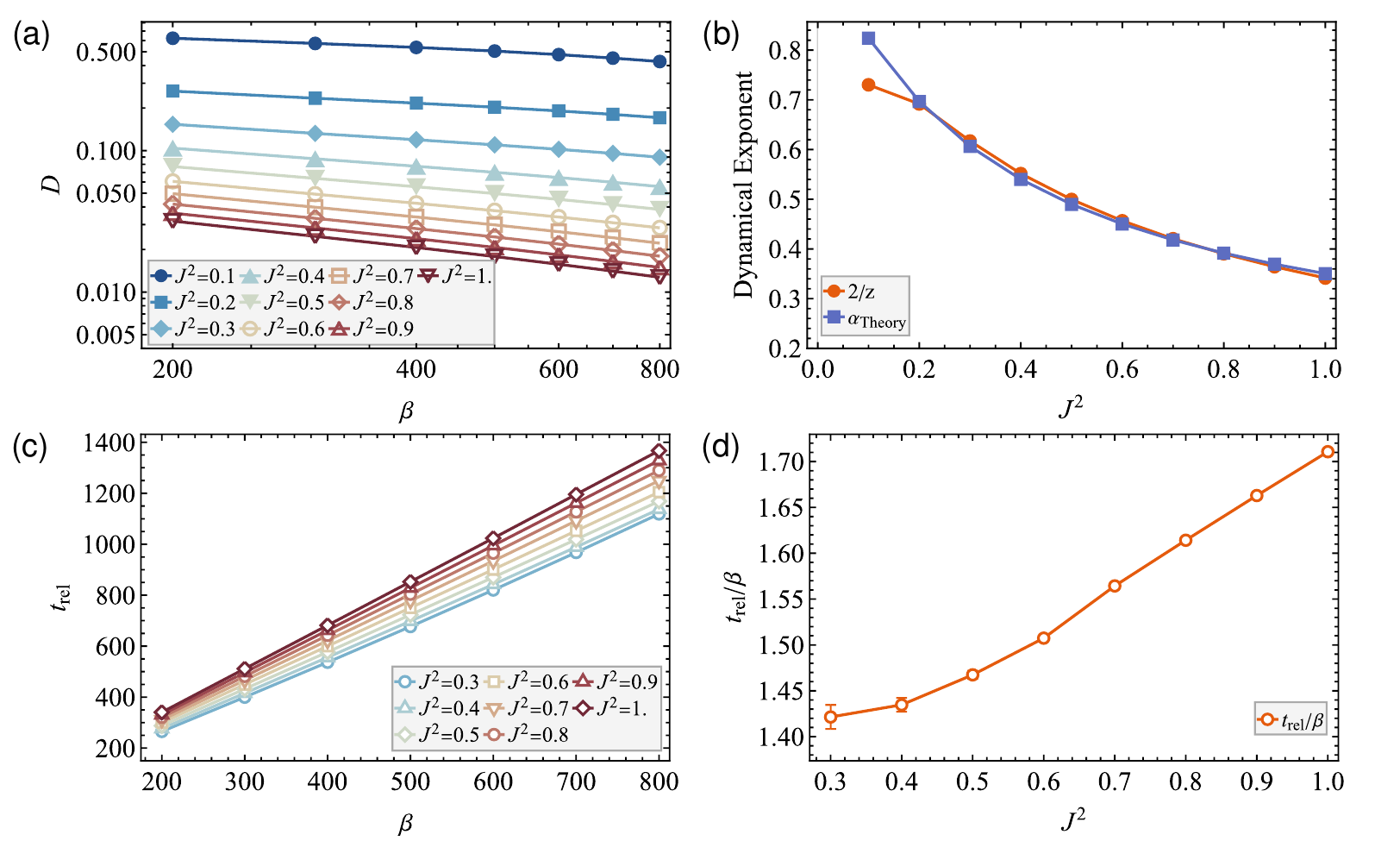}
\caption{Numerical results in the low-temperature region. All data in the figure are extrapolated to $\eta \to 0$ using different $\eta=[0.0005,0.0006,0.0007]$ in the numerics. (a) The diffusion constant is obtained by fitting the slope of the late-time correlator $\Re \Delta_<(t) \sim 2Dt + \text{const}$. The linearity in the log-log plot indicates the power-law $D \sim \beta^{2/z-1}$, where $z$ is the dynamical exponent derived from dimensional analysis. (b) The comparison between the fitted dynamical exponent $2/z$ and $\alpha_{\text{Theory}}$ from IR analysis in Eq.~\eqref{eq:JAlpha}. (c) Numerically obtained relaxation time $\rel$ by fitting the slope of $\log\left| \frac{\diff^2}{\diff t^2}\Re \Delta_<(t)\right|$ against $t$, in the time region $t\in[1.4\beta, 1.55\beta]$. The $\rel$ is plotted against $\beta$, which shows the linear behavior $\rel \sim \beta$. (d) The coefficient of the relaxation time is obtained by fitting the slope of $\rel$ against $\beta$.}
\label{fig:lowT_real_Diffusion}
\end{figure}

Here we discuss the low-temperature parameters, relying heavily on the emergent scaling symmetry at low temperature. In Fig.~\ref{fig:lowT_real_Diffusion}, we show the numerics of real-time dynamics in the low-temperature region. We focus on two quantities to describe the late-time behavior.
\begin{itemize}
    \item The diffusion constant $D$ is obtained by fitting the slope of the late-time correlator $\Re \Delta_<(t) \sim 2Dt + \text{const}$. The fitting region is chosen such that $\frac{\diff^2}{\diff t^2}\Delta_<(t)$ is small enough, ensuring that the system has entered the diffusion region.
    \item The relaxation times $\rel$ are obtained by fitting the slope of $\log\left| \frac{\diff^2}{\diff t^2}\Re \Delta_<(t) \right| \sim t$ \ \ \ (or $\log\left| \frac{\diff^2}{\diff t^2}\Im \Delta_<(t) \right| \sim t$), as illustrated in the insets in Fig.~\ref{fig:highandlowT_numerics}(c1, c2) and (d1, d2). We can exactly prove that real and imaginary part fitting leads to the same relaxation time $\rel$ in the high-temperature case, as indicated by Eq.~\eqref{eq:Delta><ansatz}. However, in low-temperature case, the numerical extraction of $\rel$ for the real and imaginary parts is slightly different, which we attribute to numerical limitations. We find that both methods give a relaxation time that scales linearly with $\beta$. Therefore, we extract $\rel$ from the slope fitting of $\log\left| \frac{\diff^2}{\diff t^2}\Re \Delta_<(t) \right| \sim t$.
\end{itemize}

Remarkably, our model exhibits power-law scaling with $D \sim \beta^{2/z-1}$, as shown in Fig.~\ref{fig:lowT_real_Diffusion}(a). Here we use the exponent $2/z-1$ to label the slope on the $D, \beta$ log-log plot, where $z$ is the dynamical exponent. This scaling arises from dimensional analysis of the diffusion behavior $\Re \Delta_<(t) \sim 2Dt + \cdots$, where $[\Delta_<(t)] \sim [\langle x^2 \rangle] \sim [t]^{2/z}$, and therefore $[D] \sim [t]^{2/z-1} \sim [\beta]^{2/z-1}$. The brackets, $[]$, indicate the unit of the physical quantity.
We observe that the dynamical exponent is continuously tunable and depends on $J$. In Fig.~\ref{fig:lowT_real_Diffusion}(b), the fitted exponent $2/z$ from the diffusion constant generally shows quantitative agreement with $\alpha_{\text{Theory}}$ from the IR analysis in Eq.~\eqref{eq:JAlpha}. When $J^2 \leq 0.1$, the error increases. This artificial numerical effect, also encountered in the high-temperature region, occurs because smaller $J$ leads to larger $t_{\rel}$, causing the system to potentially not enter the diffusion region during the fitting range $t \in [0, 1/\eta]$, which results in a larger numerical error. In conclusion, in low-temperature regions, the dynamical exponent is precisely determined by the IR theory $\alpha_{\text{Theory}}$ with $z = 2/\alpha_{\text{Theory}}$.

We are also interested in the relaxation time $\rel$ to the diffusion region in the low-temperature system. In the numerical data of Fig.~\ref{fig:lowT_real_Diffusion}(c), $\rel$ is obtained by linearly fitting $\log\left| \frac{\diff^2}{\diff t^2}\Re \Delta_<(t)\right|$ in the time region $t\in [1.4\beta, 1.55\beta]$, with an R-square value $R^2>0.999$. The fitting regions ensure that the fast decaying mode has already vanished and only one late-time decaying mode exists. Furthermore, we verify the linear behavior of $\rel \sim \beta$ by checking the R-square value $R^2>0.9998$ for all different $J^2$ in Fig.~\ref{fig:lowT_real_Diffusion}(c).
We can extract the coefficient $\rel/\beta$ (and an analogous quantity from fitting the imaginary part) for different $J^2$. The relaxation time is an $O(1)$ value as a function of $J^2$. Both $\rel/\beta$ and the analogous quantity extracted from the imaginary part show similar behavior regarding $J^2$, while they differ by a constant value. With a larger temporal fitting window, we expect the time constants of the exponential decay of both the real and imaginary parts to converge. 

We can unify the behavior of these two quantities, $\rel$ and $D$, at low temperatures within one framework. In the low temperature regime, the largest characteristic time scale is $\beta$. Therefore, we expect the relaxation time to be proportional to $\beta$ since it has the unit of time. By virtue of the dynamical exponent relating space and time scaling, the diffusion constant has the unit of $[\text{time}]^{2/z-1}$, and therefore we expect that $D \sim \beta^{\alpha - 1}$ using the relation $2/z=\alpha$.



\subsubsection{Theory of High-Temperature Parameters}

Here we discuss the high-temperature parameters, relying on some useful approximations to give an analytical treatment. At high temperature, the approximation for $\Delta_<(t) \approx \Delta_>(t) \approx \frac{t^2}{m \beta}$ works at a relatively large time scale $t\in [\beta, \rel]$, where the imaginary part of $\Delta_{\gtrless}(t)$ can be ignored due to the condition $|\iu t/m |\ll \frac{t^2}{m \beta}$ when $t\gg \beta$. We obtain $\Sigma_K$ by referring to the self-energy formula in Table \ref{tab:realtime_selfE},
\begin{equation}\label{eq:selfE_K_t}
    \Sigma_K(t) = \Sigma_<(t) + \Sigma_>(t)  = \frac{8 \iu \beta  J^2 m \left(t^2-l^2 \beta  m\right)}{\left(l^2\beta  m+t^2\right)^3}.
\end{equation}
After Fourier transformation, the corresponding $\Sigma_K(\omega)$ is
\begin{equation}
    \Sigma_K(\omega) = -\frac{2 i \pi  J^2 e^{-l \omega  \sqrt{\beta  m}} \left(l \omega  \left(\beta  l m \omega +\sqrt{\beta  m}\right)+1\right)}{l^3 \sqrt{\beta  m}}.
\end{equation}
In the small $\omega$ region, it can be expanded as the following equation
\begin{equation}
    \Sigma_K(\omega) = -\frac{2 i \pi  J^2}{l^3 \sqrt{\beta  m}}-\frac{i \pi  \beta J^2 m \omega ^2}{l \sqrt{\beta  m}} + O(\omega^3).
\end{equation}
We aim to obtain the relaxation time, so we consider the time scale when $| \omega | ^{2} \ll \left(\frac{1}{l^2\beta m} \right)$, or effectively $t^2\gg l^2\beta m$. We can approximate $\Sigma_K$ as a constant by dropping the quadratic term in $\omega$,
\begin{equation}
    \begin{split}
        &\Sigma_K(\omega) \approx - \frac{2 i \pi  J^2 }{l^3\sqrt{\beta  m}}. \\
    \end{split}
\end{equation}

By the fluctuation-dissipation theorem in Table~\ref{tab:realtime_selfE}, we obtain the retarded self-energy term,
\begin{equation}
    \Im \Sigma_R(\omega) = \frac{1}{2 \iu} \Sigma_K(\omega) \tanh{\frac{\beta \omega}{2}} \approx - \frac{\beta \omega}{2} \frac{ \pi  J^2 }{l^3\sqrt{\beta  m}},
\end{equation}
where we approximate $\tanh(\frac{\beta \omega}{2})$ as $\frac{\beta\omega}{2}$ in the high temperature limit. In principle, the real part of $\Sigma_R$ should be determined from the Kramers-Kronig relation $\operatorname{Re} \Sigma_R(\omega) = \frac{1}{\pi} \mathcal{P}\int \frac{\operatorname{Im} \Sigma_R(\omega')}{\omega' - \omega} \diff \omega'$, but the analytic continuation can be read off as
\begin{equation}
    \Sigma_R(\omega) \approx - \iu\omega \frac{\beta}{2} \frac{ \pi  J^2 }{l^3\sqrt{\beta  m}}.
\end{equation}
Notice that $\Sigma_R(\omega)\propto \omega$ also automatically satisfies the condition $\Sigma_R(\omega=0)=0$ shown in Table~\ref{tab:realtime_selfE}.

With all the approximations above, the retarded component of the Schwinger-Dyson equation in Eq.~\eqref{eq:SD_R_Delta} can be further simplified to
\begin{equation}\label{eq:highT_SD_R_omega}
    \Big(m(\omega+ \iu  \eta)^2 - \epsilon \Big)  \Delta_R(\omega) + \iu\omega  \frac{ \pi J^2 \sqrt{\beta}}{2l^3\sqrt{m}}  \Delta_R(\omega) =-2\iu ,\ \ \omega\neq 0 .
\end{equation}
In the time domain, we set $\epsilon$ to zero and Eq.~\eqref{eq:highT_SD_R_omega} becomes
\begin{equation}\label{eq:highT_SD_R_t}
    -m \frac{\diff^2 \Delta_R(t)}{\diff t^2}  - \frac{ \pi J^2 \sqrt{\beta}}{2l^3\sqrt{m}}  \frac{\diff \Delta_R(t)}{\diff t} = -2\iu \delta(t).
\end{equation}
We solve the ordinary differential equation Eq.~\eqref{eq:highT_SD_R_t} with the boundary condition $\Delta_R(0^+)=0$, $\Delta_R'(0^+)=\frac{2\iu}{m}$, $\Delta_R(0^-)=0$, $\Delta_R'(0^-)=0$. Here $\Delta_R(0^+)=0$, $\Delta_R'(0^+)=\frac{2\iu}{m}$ can be directly obtained from the free theory $J=0$. The solution for the retarded Green's function reads
\begin{equation}
    \Delta_R(t) = \iu\Theta (t) \frac{2\rel}{m}  \left(1-e^{-\frac{|t|}{\rel}}\right)\label{eq:DeltaR_highT_relax},
\end{equation}
where the relaxation time is
\begin{equation}
    \rel = \frac{2 m^{3/2} l^3}{\pi \beta^{1/2} J^2}.
\end{equation}
Similarly, the advanced component can be obtained as
\begin{equation}
    \Delta_A(t) = \iu\Theta (-t) \frac{2\rel}{m}  \left(1-e^{-\frac{|t|}{\rel}}\right)\label{eq:DeltaA_highT_relax}.
\end{equation}

Next, $\Delta_K(t)$ can be evaluated using the FDT,
\begin{equation}
    \tanh{\frac{\beta \omega}{2}} \Delta_K(\omega) \approx \frac{\beta \omega}{2} \Delta_K(\omega) = \left(\Delta_R(\omega) - \Delta_A(\omega) \right) .
\end{equation}
Using the high-temperature approximation, we can further replace $\omega$ as $\iu \partial_t$ and obtain $\Delta_K(t)$ from a first-order ordinary differential equation, 
\begin{equation*}
    \frac{\diff \Delta_K(t)}{\diff t}=\frac{2}{\iu\beta}(\Delta_R(t)-\Delta_A(t)).
\end{equation*}
Consequently, we obtain $\Delta_K(t)$ and then the real-part $\Delta_{\gtrless}$, which leads to the final result
\begin{align}
\Delta_K(t) &= 4(e^{-|t|/\rel}-1 + |t|/\rel)\rel^2 /(m\beta) \label{eq:DeltaKansatz}\\
\Delta_{\gtrless}(t) &= 2(e^{-|t|/\rel}-1 + |t|/\rel)\rel^2 /(m\beta)  \pm \iu \sgn(t) \frac{\rel}{m}  \left(1-e^{-\frac{|t|}{\rel}}\right)\label{eq:Delta><ansatz}.
\end{align}
From this result, the system displays diffusive behavior at late time with $\Delta_<(t) \approx 2Dt + C$, with constant $C=-2\rel^2/(m\beta)$. The diffusion constant $D$ can be simply related to the $\rel$ by
\begin{equation}\label{eq:highT_diff_const}
    D = \rel/m\beta =  \frac{2 m^2 l^3}{\pi \beta^{3/2} J^2} .\\
\end{equation}

\begin{figure}[t]
\centering
\includegraphics[scale=0.55]{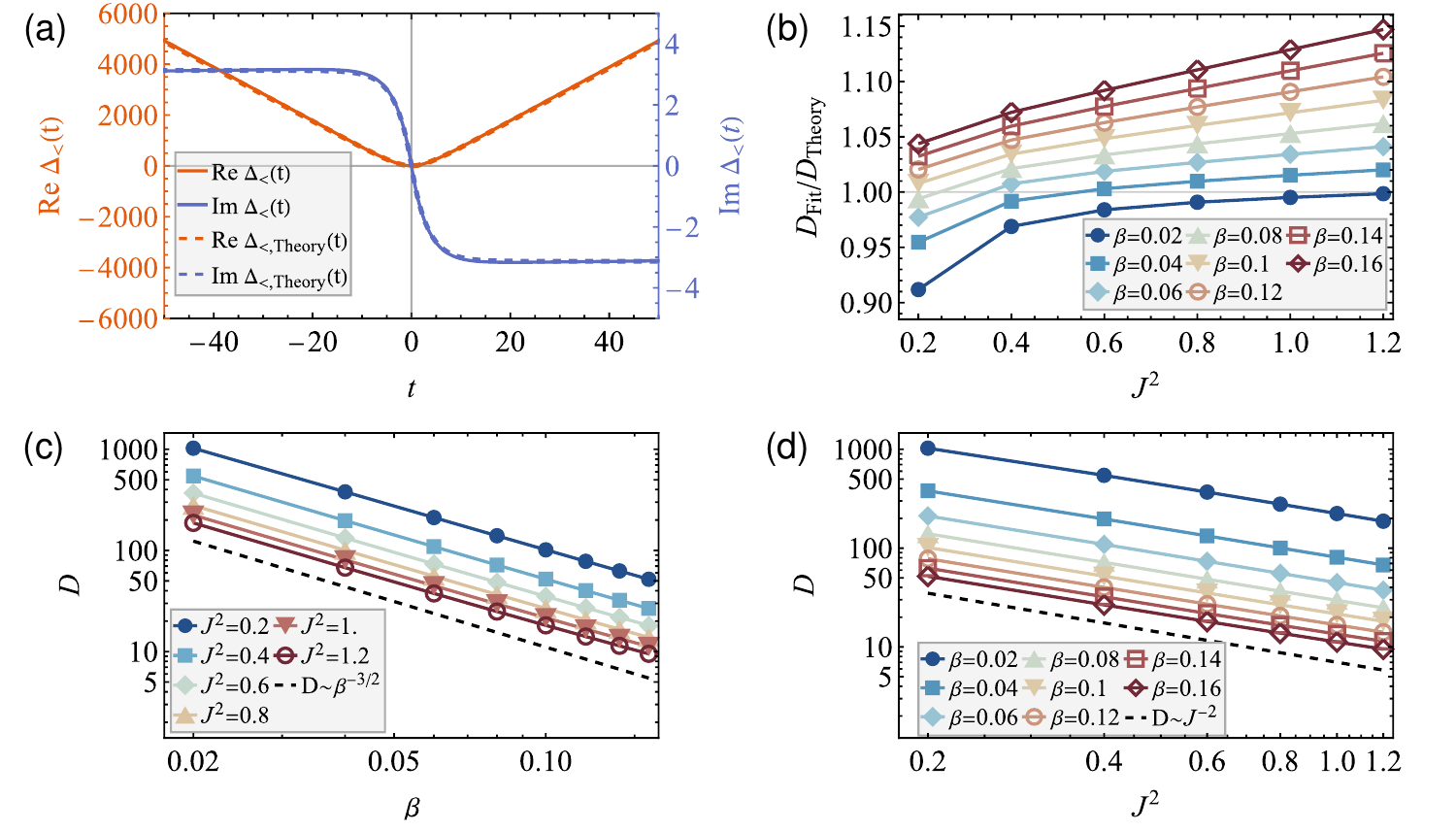}
\caption{Numerical results in the high-temperature region. (a) The typical dynamics of the correlator $\Delta_<(t)$, with $\beta=0.06, J^2=0.8, \eta=0.01$. We plot $\Re \Delta_<(t)$ and $\Im \Delta_<(t)$ on different scales, as separately labeled by the left and right axes. The red line and blue line represent $\Re \Delta_<(t)$ and $\Im \Delta_<(t)$, respectively. For each color, the solid line represents the numerics, and the dashed line represents the theoretical prediction Eq.~\eqref{eq:Delta><ansatz}, with $\rel=3.1278$. (b) The ratio between the numerically fitted diffusion constant and the theoretically predicted diffusion constant in Eq.~\eqref{eq:highT_diff_const}. (c) The numerically obtained diffusion constant plotted against $\beta$ in a log-log plot. We provide the black dashed line of $D \sim \beta^{-3/2}$ scaling, which serves as a guide for the eye. (d) The numerically obtained diffusion constant plotted against $J^2$ in a log-log plot. We provide the black dashed line of $D \sim J^{-2}$ scaling, which also serves as a guide for the eye. In Fig.~(b, c, d), the results have been extrapolated to $\eta \to 0$ using different $\eta=[0.007,0.01,0.015]$.}
\label{fig:highT_real_Diffusion}
\end{figure}

To benchmark our theory, we directly perform numerical calculations based on the Schwinger-Dyson equation. In Fig.~\ref{fig:highT_real_Diffusion}, we provide concrete evidence for our high-temperature solution.

Firstly, we can check that the high-temperature theory, Eq.~\eqref{eq:Delta><ansatz}, perfectly matches the numerics, as indicated by the overlap between the dashed lines and solid lines in Fig.~\ref{fig:highT_real_Diffusion}. Additionally, in the derivation of the self-energy, Eq.~\eqref{eq:selfE_K_t}, we assume the real part of $\Delta_<(t)$ is significantly larger than the imaginary part by using the free theory solution, and the numerics show that this approximation remains true for the exact dynamics.

Secondly, after verifying the functional form of $\Delta_<(t)$, we focus on the characteristic relaxation time $\rel$, or equivalently, the diffusion constant $D$ in the high-temperature region. In Fig.~\ref{fig:highT_real_Diffusion}(b,c,d), the diffusion constant (or $\rel$) is obtained by directly fitting the formula, Eq.~\eqref{eq:Delta><ansatz}, in the reliable time region $t \in [0, 1/\eta]$. We choose different cutoffs, $\eta = [0.007, 0.01, 0.015]$, in the numerics and extrapolate the data to $\eta = 0$.

In Fig.~\ref{fig:highT_real_Diffusion}(b), we quantitatively verify the numerically obtained diffusion constant against the theoretically predicted value in Eq.~\eqref{eq:highT_diff_const}. In the small $\beta \to 0$ limit, the ratio $D_{\text{Fit}}/D_{\text{Theory}} \approx 1$ within an error of 5\%. We note that the error increases as $J$ approaches zero. This is an artificial numerical effect because smaller $J$ leads to larger $t_{\rel}$, and the system might not enter the diffusion region during the fitting region $t \in [0, 1/\eta]$, which causes larger numerical error.

In Fig.~\ref{fig:highT_real_Diffusion}(c) and Fig.~\ref{fig:highT_real_Diffusion}(d), we illustrate that the diffusion constant follows a power-law dependence on $\beta$ and $J$ by observing the linearity in the log-log plot. Most importantly, we extract the scaling law by comparing with $D \sim \beta^{-3/2}$ and $D \sim J^{-2}$, which again confirms the correctness of our prediction in the high-temperature region as given by Eq.~\eqref{eq:highT_diff_const}.

\section{Lyapunov Physics}
\label{sec:lyapunov}

In this section, we setup the calculation of an appropriate out-of-time-order correlator (OTOC) \cite{larkin1969quasiclassical,almheiriApologiaFirewalls2013,Shenker_2014,robertsDiagnosingChaosUsing2015,robertsLocalizedShocks2015,shenkerStringyEffectsScrambling2015,hosurChaosQuantumChannels2016a,MSS,stanfordManybodyChaosWeak2016a,maldacenaConformalSymmetryIts2016,Maldacena_2016,zhuMeasurementManybodyChaos2016,haydenMeasuringScramblingQuantum2016,chenTunableQuantumChaos2017a,chenCompetitionChaoticNonChaotic2017,songStronglyCorrelatedMetal2017a,patelQuantumButterflyEffect2017,vonkeyserlingkOperatorHydrodynamicsOTOCs2018a,xuLocalityQuantumFluctuations2019a,zhangQuantumChaosUnitary2019a,kitaevRelationMagnitudeExponent2019,kitaevObstacleSubAdSHolography2021,zhangTwowayApproachOutoftimeorder2022,lewis-swanUnifyingScramblingThermalization2019a} in order to extract a quantum Lyapunov exponent. We then report the results of a numerical study of the derived equations. In particular, at low temperature we find a chaos exponent which is proportional to $1/\beta$ and grows with increasing $J$. We do not observe any satuaration of this growth and the chaos exponent comes close to $90\%$ of maximal at the largest $J$s we studied.


We study the 4-point function of $x$s by performing a perturbation around the saddle point of the full effective action Eq.~\eqref{eq:FullAction}
\begin{equation}\label{eq:GSigma_pert}
    \begin{split}
        G(\tau_1,\tau_2) = \tilde{G}(\tau_1,\tau_2) + \delta G(\tau_1,\tau_2)  \\
        \Sigma(\tau_1,\tau_2) = \tilde{\Sigma}(\tau_1,\tau_2) + \delta \Sigma(\tau_1,\tau_2). \\
    \end{split}
\end{equation}
Then the effective action becomes
\begin{equation}\label{eq:Action_deltaGdeltaSigma}
    \begin{split}
        \frac{I_{\text{eff}}[\delta G,\delta \Sigma]}{N} =& \frac{1}{4} \int \diff{\tau_1}\diff{\tau_2}\diff{\tau_3}\diff{\tau_4} \delta \Sigma(\tau_1, \tau_2) \left( \tilde{G}(\tau_1, \tau_3) \tilde{G}(\tau_2, \tau_4)  \right) \delta \Sigma(\tau_3, \tau_4) + \frac{1}{2} \int \diff{\tau_1}\diff{\tau_2} \\
        & \Bigg[\delta G(\tau_1,\tau_2)\delta \Sigma(\tau_1,\tau_2) 
        + (\delta G(\tau_1,\tau_2) )^2 \frac{f''(\tilde{\Delta})}{2} (2(\delta(\tau_1-\tau_2)-1))^2 \partial_{\tau_1}\partial_{\tau_2}\tilde{G}(\tau_1,\tau_2) \\ 
        &+ (\delta G(\tau_1,\tau_2) ) f'(\tilde{\Delta}) (2(\delta(\tau_1-\tau_2)-1)) \partial_{\tau_1}\partial_{\tau_2}(\delta G(\tau_1,\tau_2) ) \Bigg] \\
    \end{split}
\end{equation}
After integrating out $\delta \Sigma$, the effective action reads
\begin{equation}\label{eq:Action_deltaG}
    \frac{I_{\text{eff}}[\delta G]}{N} = \int \diff{\tau_1}\diff{\tau_2}\diff{\tau_3}\diff{\tau_4} \delta G(\tau_1,\tau_2) \left(\frac{1}{4}K^{-1} - \frac{1}{2}(\mathrm{I} \circ \Lambda) \right) \delta G(\tau_3,\tau_4),
\end{equation}
where the kernel $K$ is
\begin{equation}
    K(\tau_1,\tau_2,\tau_3,\tau_4) = -  \tilde{G}(\tau_1, \tau_3) \tilde{G}(\tau_2, \tau_4) ,
\end{equation}
the interaction 2-point vertex induced by the magnetic field is
\begin{equation}
    \begin{split}
        \Lambda(\tau_{1},\tau_{2}) &=  \Bigg[ \frac{f''(\tilde{\Delta})}{2} (2(\delta(\tau_1-\tau_2)-1))^2 \partial_{\tau_1}\partial_{\tau_2}\tilde{G}(\tau_1,\tau_2) +  f'(\tilde{\Delta}) (2(\delta(\tau_1-\tau_2)-1)) \left(\partial_{\tau_1}\partial_{\tau_2} \right) \Bigg], \\
    \end{split}
\end{equation}
and the identity 4-point vertex for the boson is
\begin{equation}
    \mathrm{I}(\tau_1,\tau_2,\tau_3,\tau_4) = \frac{1}{2}\left( \delta(\tau_1-\tau_3)\delta(\tau_2-\tau_4) + \delta(\tau_1-\tau_4)\delta(\tau_2-\tau_3)\right).
\end{equation}
We can check that $\int \diff{\tau_1}\diff{\tau_2}\diff{\tau_3}\diff{\tau_4} \delta G(\tau_1,\tau_2) \left( \frac{1}{2}(\mathrm{I}\circ \Lambda) \right) \delta G(\tau_3,\tau_4)$ is the same magnetic interaction term in Eq.~\eqref{eq:Action_deltaGdeltaSigma}.

The 4-point function of $x$ can be regarded as the 2-point function of the perturbed Green's function in the effective action Eq.~\eqref{eq:Action_deltaG}. We introduce
\begin{equation}\label{eq:Lyap_F_calc}
    \begin{split}
        F(\tau_1,\tau_2,\tau_3,\tau_4) &= \langle \delta G(\tau_1,\tau_2)  \delta G(\tau_3,\tau_4) \rangle \\
        &= \frac{1}{2} \left(\mathrm{I} - 2K \circ (\mathrm{I} \circ \Lambda))\right)^{-1} K\circ \mathrm{I} \frac{1}{N}    \\
    \end{split} 
\end{equation}
Then we can directly prove that the 4-point correlation function is generated by a series of ladder diagrams. To simplify the above equation we notice that 
\begin{equation}\label{eq:Lyap_KI}
    \begin{split}
        (K \circ \mathrm{I})(\tau_1,\tau_2,\tau_5,\tau_6) &= \int \diff{\tau_3}\diff{\tau_4}K(\tau_1,\tau_2,\tau_3,\tau_4) \mathrm{I}(\tau_3,\tau_4,\tau_5,\tau_6) \\
        &= - \frac{1}{2} \tilde{G}(\tau_1, \tau_5) \tilde{G}(\tau_2, \tau_6)  - \frac{1}{2} \tilde{G}(\tau_1, \tau_6) \tilde{G}(\tau_2, \tau_5) \\
    \end{split}
\end{equation}
where we can further define the decoupled 4-point function
\begin{equation}\label{eq:Lyap_F0}
    F_0(\tau_1,\tau_2,\tau_3,\tau_4)=-\frac{1}{N}\left(\tilde{G}(\tau_1,\tau_4)\tilde{G}(\tau_2,\tau_3)+\tilde{G}(\tau_1,\tau_3)\tilde{G}(\tau_2,\tau_4) \right).
\end{equation}
Combining all the results in Eq.~\eqref{eq:Lyap_F_calc}, \eqref{eq:Lyap_F0}, \eqref{eq:Lyap_KI} we get the final result
\begin{equation}
    \begin{split}
        F &= \left(\mathrm{I} - 2 K \circ (\mathrm{I} \circ \Lambda))\right)^{-1} F_0 \\
        &= \sum_{n=0} \left(2 K \circ (\mathrm{I} \circ \Lambda)) \right)^n F_0 \\
    \end{split}
\end{equation}
Therefore $2 K \circ (\mathrm{I} \circ \Lambda))$ is the effective ladder diagram in our model, which reads 
\begin{equation}\label{eq:Klad_imag}
    \begin{split}
        K_{\text{lad}}(\tau_1,\tau_2,\tau_3,\tau_4) &\equiv  \left(2 K \circ (\mathrm{I} \circ \Lambda)(\tau_1,\tau_2,\tau_3,\tau_4)\right)  \\
        &= -2\tilde{G}(\tau_1, \tau_3) \tilde{G}(\tau_2, \tau_4) 
        \Bigg[(\frac{f''(\tilde{\Delta})}{2} (2(\delta(\tau_3-\tau_4)-1))^2 \partial_{\tau_3}\partial_{\tau_4}\tilde{G}(\tau_3,\tau_4) \\
    &\qquad\qquad+ (f'(\tilde{\Delta}) (2(\delta(\tau_3-\tau_4)-1)) \left(\partial_{\tau_3}\partial_{\tau_4} \right) \Bigg] .\\
    \end{split}
\end{equation}

Finally, we consider the OTOC as a special kind of 4-point correlator on the double Keldysh contour with the parameterization.
\begin{align}
        &\text{OTOC}(t_1,t_2,t_3,t_4) = \frac{1}{N} F(\tau_1,\tau_2,\tau_3,\tau_4), \\
        &\text{with}\ \tau_1=\beta/2+\iu t_1,\tau_2=\iu t_2,\tau_3=\beta/4+t_3,\tau_4=-\beta/4+t_4 \\
\end{align}
It is known that any 4-point function satisfies the Bethe-Salpeter equation even for the deformed contour \cite{kitaevRelationMagnitudeExponent2019,kitaevObstacleSubAdSHolography2021,zhangTwowayApproachOutoftimeorder2022},
\begin{equation}\label{eq:Bethe_Salpeter}
    F(\tau_1,\tau_2,\tau_3,\tau_4) = \int_{\mathcal{C}} \diff{\tau_5} \diff{\tau_6} K_{\text{lad}}(\tau_1,\tau_2,\tau_5,\tau_6)F(\tau_5,\tau_6,\tau_3,\tau_4),
\end{equation}
\begin{figure}[hbt]
\centering
\begin{tabular}{c}
\begin{tikzpicture}[scale=0.5,baseline={([yshift=-7pt]current bounding box.center)}]
\draw [->,>=stealth] (-50pt,-92pt) -- (200pt,-92pt) node[right]{\scriptsize  $\Im t$};
\draw [->,>=stealth] (0pt, -135pt) -- (0pt,40pt) node[right]{\scriptsize  $\Re t$};
\draw[thick,gray] (0pt,20pt)--(0pt,-20pt);
\draw[thick,far arrow,gray] (140pt,-20pt)--(0pt,-20pt);
\draw[thick,far arrow,gray] (0pt,-24pt)--(140pt,-24pt);
\draw[thick,gray] (0pt,-24pt)--(0pt,-90pt);
\draw[<-,>=stealth, thick,gray] (0pt,10pt)--(0pt,-20pt); 
\filldraw (140pt,-22pt) circle (2pt) node[right]{\scriptsize$\tau_1 = \iu t+\beta/2$};
\filldraw[blue] (60pt,-20pt) circle (2pt)node[above right]{\rom{1}};
\filldraw[blue] (60pt,-24pt) circle (2pt)node[below right]{\rom{2}, $\tau_5$};
\filldraw[blue] (60pt,-90pt) circle (2pt)node[above right]{\rom{3}};
\filldraw[blue] (60pt,-94pt) circle (2pt)node[below right]{\rom{4}, $\tau_6$};

\draw[thick,far arrow,gray] (140pt,-90pt)--(0pt,-90pt);
\draw[thick,far arrow,gray] (0pt,-94pt)--(140pt,-94pt);
\draw[far arrow, thick,gray] (0pt,-130pt)--(0pt,-94pt); 
\filldraw (140pt,-92pt) circle (2pt) node[below right]{\scriptsize$\tau_2 = \iu t$};
\filldraw[black] (0pt,-57pt) circle (2pt) node[left]{\scriptsize $\tau_3 = \beta
/4$};
\filldraw[black] (0pt,-127pt) circle (2pt) node[left]{\scriptsize $\tau_4 = -\beta
/4$};
\end{tikzpicture}
\end{tabular}
\caption{The path-integral contour $\mathcal{C}$ for the out-of-time-ordered 4-point correlation function OTOC$(t,t,0,0)$. The blue points denote the contribution from $\tau_5,\tau_6$ in the kernel.
}
\label{figcontourfinite}
\end{figure}
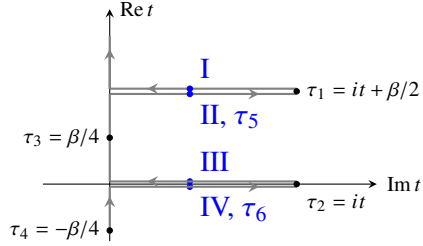
where we ignore the inhomogeneous term $F_0$, which means the ladder with no rungs, since it's much smaller than the OTOC contribution. In Eq.~\eqref{eq:Bethe_Salpeter}, $\mathcal{C}$ refers to the deformed double Keldysh contour. 
Following the same technique used in the analysis of SYK, the imaginary variable $\tau_5$ can be represented as $\iu t_5 + \beta/2$ and $\iu t_5$, with each single $t_5$ having two points in the backward and forward contour. These points are denoted by the blue dots \rom{1}, \rom{2}, \rom{3}, and \rom{4}, respectively. For $\tau_1=\iu t + \beta/2$, the contributions of blue points \rom{3} and \rom{4} cancel out, but the contribution from points \rom{1} and \rom{2} is non-vanishing, leading to the contribution $\Theta(t_1-t_5)(\iu G_<(t_{1},t_{5})-\iu G_>(t_{1},t_{5}))=-\iu G_R(t_1, t_5)$ \cite{kitaevRelationMagnitudeExponent2019,kitaevObstacleSubAdSHolography2021,zhangTwowayApproachOutoftimeorder2022}. Similarly, we find that $\tau_6$ only contributes from blue points \rom{3} and \rom{4}, leading to $-\iu G_R(t_2, t_6)$. This automatically leads to the simplification that $\tau_5$ and $\tau_6$ differ by an imaginary time $\beta/2$, which means the delta function in Eq.~\eqref{eq:Klad_imag} has no contribution. Additionally, $F(\tau_5,\tau_6,\tau_3,\tau_4)$ will be invariant when considering points \rom{1} and \rom{2}, since the causal relation between the blue points and $\tau_3$ and $\tau_4$ is fixed with certain imaginary time. After considering these structures, the real-time kernel reads.
\begin{equation}
    \begin{split}
        K_{\text{lad},R}(t_1,t_2,t_5,t_6) &= -2\tilde{G}_R(t_1,t_3) \tilde{G}_R(t_2, t_5) 
        \Bigg[f''(\tilde{\Delta}_W(t_5,t_6))  \partial_{t_5}\partial_{t_6}\tilde{G}_W(t_5,t_6) + 2 f'(\tilde{\Delta}_W(t_5,t_6))  \left(\partial_{t_5}\partial_{t_6} \right) \Bigg] \\
    \end{split}
\end{equation}
where the Wightman Green's functions are defined by
\begin{equation}
    \iu G_W(t) \equiv \langle x(t-\iu \beta/2) x(0) \rangle,\ \ \iu \Delta_W(t) \equiv \langle \left(x(t-\iu \beta/2)- x(0) \right)^2 \rangle
\end{equation}
and their relation to the spectral function can be found in the Appendix~\ref{app:analyticalJ0}. 
Finally, we take an ansatz for OTOC
\begin{equation}
    \mathrm{OTOC}(t_1, t_2, 0, 0) = e^{\varkappa (t_1+t_2)/2} \mathcal{F}(t_1-t_2),
\end{equation}
and plug it into the Bethe-Salpeter equation 
\begin{equation}
    \text{OTOC}(t_1,t_2,0,0) = \int_{\mathbb{R}} \diff{t_5} \diff{t_6} K_{\text{lad},R}(t_1,t_2,t_5,t_6)\text{OTOC}(t_5,t_6,0,0).
\end{equation}
These manipulations yield at last an eigenvalue problem for the OTOC,
\begin{equation}\label{eq:OTOC_eigen}
    \boxed{\mathcal{F}(\omega) = -2\int \frac{\diff \omega'}{2\pi} \left|G_R(\omega+\iu \frac{\varkappa}{2})\right|^2 \left( \Lambda_{W,1}(\omega-\omega') + \Lambda_{W,2}(\omega-\omega') \left( \frac{\varkappa^2}{4}+(\omega')^2\right)\right) \mathcal{F}(\omega')},
\end{equation}
where the Wightman rung diagrams are
\begin{equation}
    \begin{split}
        \Lambda_{W,1}(\omega) & = \int \diff{t} f''(\tilde{\Delta}_W(t)) (-\partial_t^2 \tilde{\Delta}_W(t)) e^{\iu \omega t}\\
        \Lambda_{W,2}(\omega) &= 2 \int \diff{t}  f'(\tilde{\Delta}_W(t)) e^{\iu \omega t}. \\
    \end{split}
\end{equation}
Additionally, the analytical continuation to $G_R(\omega+\iu \frac{\varkappa}{2})$ can be represented using spectral function, which reads 
\begin{equation}
    G_R(\omega + \iu \varkappa/2) =\frac{1}{-2\iu} \int \diff{\omega'} \frac{\rho_{\Delta}(\omega')}{\omega' - (\omega+ \iu \varkappa/2)}.
\end{equation}

\begin{figure}[t]
\centering
\includegraphics[scale=0.5]{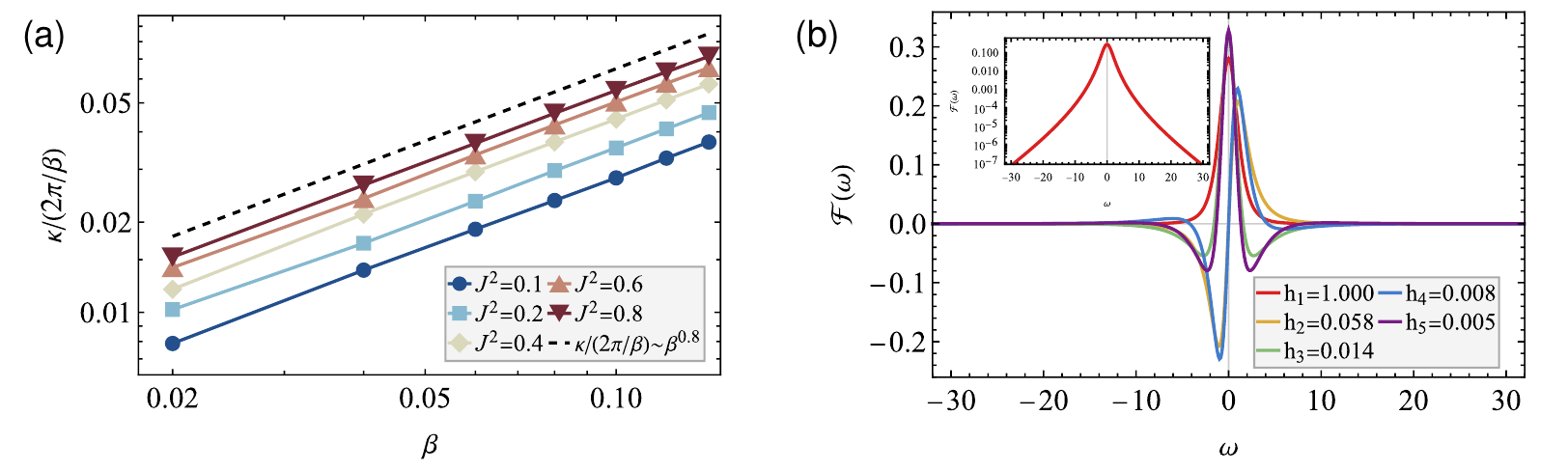}
\caption{The Lyapunov physics in the high-temperature region. All the data are extrapolated to $\eta=0$ using numerics with $\eta=[0.007, 0.01, 0.015]$. (a) The Lyapunov exponent $\varkappa/(2\pi/\beta)$ exhibits a power law against the inverse temperature $\beta$ for different $J^2$. We provide the black dashed line of $\varkappa/(2\pi/\beta) \sim \beta^{0.8}$ scaling as a guide for the eye. (b) The eigenvalue $h_i$ and the corresponding eigenfunction in the frequency domain $\mathcal{F}_i(\omega)$ are obtained by solving Eq.~\eqref{eq:OTOC_eigen} for a typical high-temperature setting, where $\beta=0.06$ and $J^2=0.6$. In the numerics, we take $\eta=0.007$, and the corresponding Lyapunov exponent is $\varkappa/(2\pi/\beta)\approx 0.03317$. $h_1,\cdots, h_5$ are the first 5 maximal eigenvalues. The inset is the log plot of the eigenfunction corresponding to the maximal eigenvalue $\mathcal{F}_1(\omega)$.}
\label{fig:Lyapunov_highT}
\end{figure}

We can diagonalize the RHS of Eq.~\eqref{eq:OTOC_eigen} to find the largest eigenvalue. As a sanity check of the numerics, we find that all the eigenvalues are real. We first discuss the numerical results for the high-temperature Lyapunov physics in Fig.~\ref{fig:Lyapunov_highT}. We find the Lyapunov exponent $\varkappa/(2\pi/\beta)$ exhibits the same power law for all the $J^2$ in Fig.~\ref{fig:Lyapunov_highT}(a). Via linear fitting on a log-log plot, we extract the power-law behavior $\varkappa/(2\pi/\beta) \sim \beta^{0.8}$. In the high-temperature region, the Lyapunov exponent is far from the chaos bound, i.e., $\varkappa/(2\pi/\beta) \ll 1$. We also provide the eigenvalue $h_i$ and the corresponding eigenfunction in the frequency domain $\mathcal{F}_i(\omega)$ by solving Eq.~\eqref{eq:OTOC_eigen} for a typical high-temperature setting, where $\beta=0.06, J^2=0.6$. As shown in Fig.~\ref{fig:Lyapunov_highT}(b), the largest eigenfunction corresponds to a positive and symmetric function centered at $\omega=0$. Other eigenfunctions are negative in some regions, and the cancellation leads to smaller eigenvalues for those modes.

\begin{figure}[t]
\centering
\includegraphics[scale=0.55]{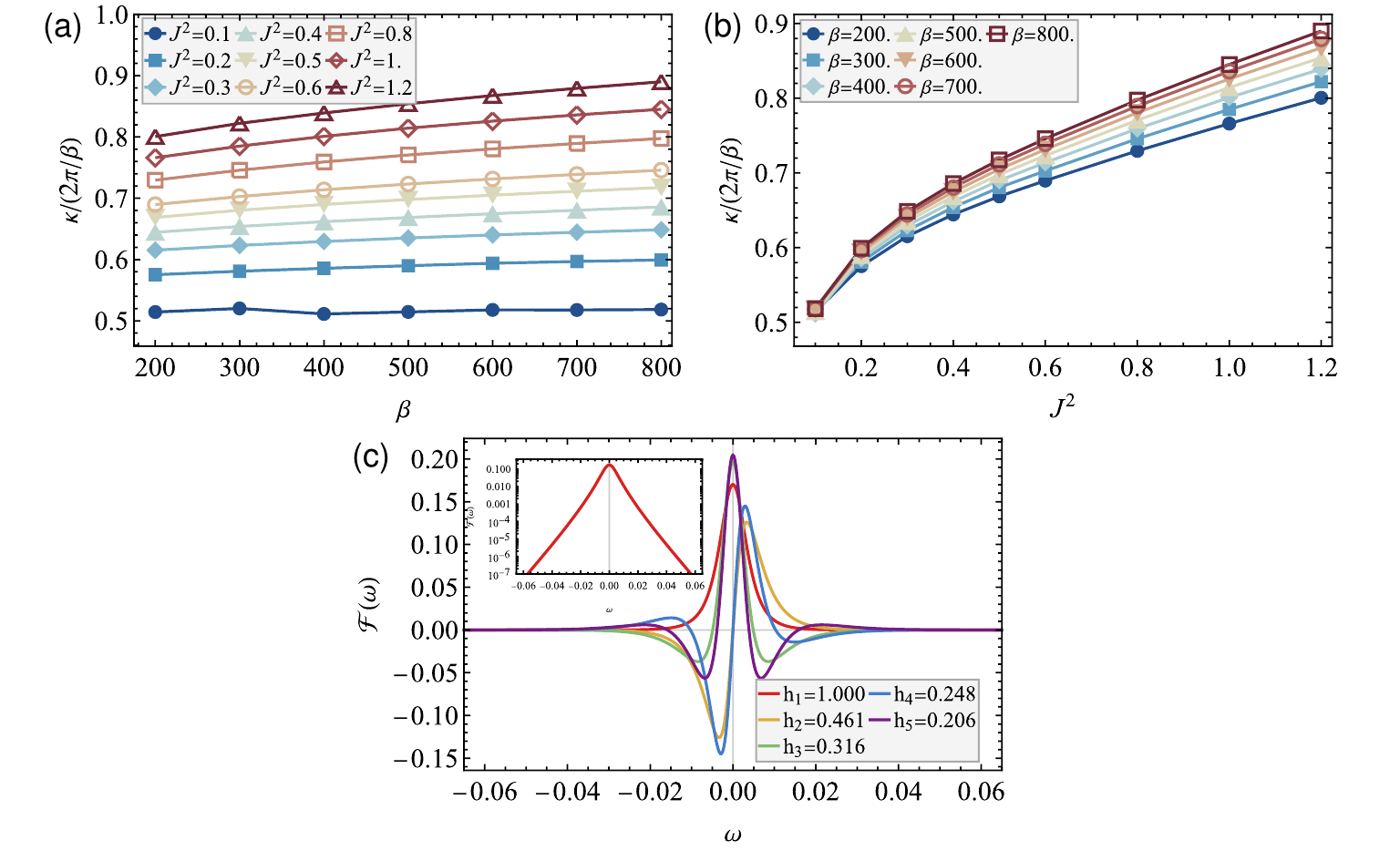}
\caption{The Lyapunov physics in the low-temperature region. All the data are extrapolated to $\eta=0$ using numerics with $\eta=[0.0005, 0.0006, 0.0007]$. (a) The Lyapunov exponent $\varkappa/(2\pi/\beta)$ only slightly depends on the inverse temperature $\beta$ for different $J^2$. (b) In the numerically available region, the Lyapunov exponent grows with increasing $J$, leaving open the possibility of a maximal chaotic region. (c) The eigenvalue $h_i$ and the corresponding eigenfunction in the frequency domain $\mathcal{F}_i(\omega)$ are obtained by solving Eq.~\eqref{eq:OTOC_eigen} for a typical low-temperature setting, where $\beta=400$ and $J^2=0.5$. In the numerics, we take $\eta=0.0005$, and the corresponding Lyapunov exponent is $\varkappa/(2\pi/\beta) \approx 0.630$. $h_1, \cdots, h_5$ are the first 5 maximal eigenvalues. The inset is the log plot of the eigenfunction corresponding to the maximal eigenvalue $\mathcal{F}_1(\omega)$.}
\label{fig:Lyapunov_lowT}
\end{figure}

We show the numerical results for the low-temperature Lyapunov physics in Fig.~\ref{fig:Lyapunov_lowT}. 
In Fig.~\ref{fig:Lyapunov_lowT}(a), the Lyapunov exponent $\varkappa$ is seen to be set by $1/\beta$, albeit with some mild temperature dependence. The smaller the $J$, the weaker the temperature dependence of $\varkappa/(2\pi/\beta)$ in this range of $\beta$. The results are all consistent with the chaos bound, $\varkappa \leq 2\pi/\beta$. In Fig.~\ref{fig:Lyapunov_lowT}(b), the Lyapunov exponent increases monotonically when $J^2$ increases. Due to numerical limitations, we are only able to probe the parameter region $\beta < 800$, $J^2 \sim O(1)$ with controlled numerical error. Hence, it is still open whether the system approaches maximal chaotic behavior at the largest $J$s and $\beta$s. But we observe no saturation with increasing $J$. Additionally, we also plot the eigenfunction $\mathcal{F}_i(\omega)$ associated with the eigenvalue in the low-temperature region. We find that the eigenfunctions $\mathcal{F}_1, \cdots, \mathcal{F}_5$ are qualitatively the same in the high-temperature and low-temperature regions. In the intermediate large $\omega$ region, $\mathcal{F}_1(\omega)$ shows exponential decaying behavior. However, the high-temperature eigenfunction in the small $\omega$ region will significantly deviate from the exponential decay, as illustrated in Fig.~\ref{fig:Lyapunov_highT}(b). But this phenomenon is not significant in Fig.~\ref{fig:Lyapunov_lowT}(c).

\section{Discussion}
\label{sec:discussion}

In this paper, we presented a systematic study of the Magnetic Maze (MM), including its Euclidean dynamics, real-time dynamics, and Lyapunov physics. The structure of the large-$N$ path integral framework, specifically the presence of certain time derivatives in the equations of motion, guarantees the absence of a glassy equilibrium state in this model. We formulated all the governing equations for general vector potential correlations, but we focused our analysis on the case where those correlations have a special power-law falloff, $f(\Delta) \sim J^2/\Delta$. We found a rather rich theory in the low-energy regime, one with tunable dynamical exponent, which gives a new kind of controllable scale invariant quantum system.

We conclude by highlighting some open questions and directions for further work:

\paragraph{Low Energy EFT} There is a lot we know about the Magnetic Maze at low temperature. We know the low-energy equations of motion have two emergent symmetries: reparameterization and rescaling. However, it is unclear what signatures these symmetries have in Lorentzian time, or what form any Goldstone bosons for these symmetries might take.

\paragraph{Maximal chaos?} We did not observe any signature that the chaos exponent saturated with increasing $J$, therefore it is possible that the Magnetic Maze approaches maximal chaos in the limit of large $J$ and low temperature. It would be very interesting to better understand this regime, possibly by making use of the Low Energy EFT.

\paragraph{Gravity dual?} It is interesting to ask if the low-energy physics of the Magnetic Maze could have a gravity dual. This question is particularly interesting if it can be shown that the Magnetic Maze is maximally chaotic at low temperature in the strong coupling limit. The dynamical exponent $z$ is a key feature to match, along with the pattern of symmetries.

\paragraph{Generalizations} There are interesting generalizations of the Magnetic Maze, including the possibility of adding random electric fields (which could be equivalent to adding random magnetic fields to the $p$-spherical model), the possibility of considering externally imposed time-dependent fields (analogous to a random circuit model), and various many-body generalizations. On the latter point, we note that the variable dynamical exponent allows one to tune the particle interactions to be relevant or marginal even in high dimensions. We can also generalize to other observables, such as the spectral form factor which probes statistical properties of the energy spectrum.

\paragraph{Finite $N$} It is also interesting to explore finite $N$ corrections to our analysis. A systematic $1/N$ expansion would be useful as well as an analysis at small finite $N$, even $N=2$. The problem of random magnetic fields is well-studied in 2d, but the special power-law correlations we studied may not have been considered before. Finally, we note that cosmic magnetic fields can have a random character \cite{PhysRevE.107.065206,Chandran_2000}, and our work might provide an alternative perspective on such systems via a $1/N$ expansion.

\acknowledgments

 We thank Xiao Chen, Ping Gao, Martin Sasieta
and Pengfei Zhang for useful discussions.
 T.-G. Z. acknowledges support from the Tsinghua Visiting Doctoral Students
 Foundation. MW acknowledges support from the Joint Quantum Institute. BGS acknowledges support from the U.S. Department of Energy through DE-SC0009986.

\appendix


\section{Uniform Field Problem and Physical Scales}
\label{app:crash_course}

In this appendix, we review the basic physical scales in the magnetic maze. Throughout the paper, we absorb the charge into the vector potential and set $k_B= 1$, but we keep $\hbar$ explicit in this appendix. It is useful to approach this analysis by first discussing the case of a uniform magnetic field. This clarifies the physical meaning of the dimensionless parameters in the more general problem.

In $N$ dimensions, a random spatially uniform magnetic field is an antisymmetric tensor $B_{ij}$ which we take to be an $N \times N$ antisymmetric matrix with matrix elements given by Gaussian random variables with zero mean and variance equal to $\Bb^2/N$. Antisymmetric matrices don't have real eigenvalues, but they do have eigenplanes in which the matrix takes the form $\begin{bmatrix} 0 & \lambda \\ - \lambda & 0 \end{bmatrix}$ for some field strength $\lambda$. It can further be shown that the field strengths, which vary from plane to plane, have a semi-circle distribution at large $N$,
\begin{equation}
    \rho(\lambda) = \frac{N}{2\pi \Bb^2}\sqrt{4 \Bb^2 - \lambda^2},
\end{equation}
for $0 < \lambda < 2 \Bb$. When $N$ is odd, there is one special direction that does not experience any magnetic field.

\subsection{Review of the 2d Uniform Field Problem}

The $N$-dimensional problem thus breaks up into $N/2$ copies of the $2$-dimensional problem. In the 2d classical problem with a uniform field $\lambda$, a particle with speed $v$ undergoes uniform circular motion with angular frequency $\omega_c = \lambda/m$ and radius $r_c = v/\omega_c$. In the corresponding 2d quantum problem, we also have the magnetic length, $\ell_B^2 \sim \frac{\hbar}{\lambda}$, and the characteristic energy of the Landau levels, $\hbar \omega_c \sim \frac{\hbar^2}{m \ell_B^2}$.

For a particle with a thermal velocity, $v \sim \sqrt{T/m}$, the ratio of the classical orbit size to the magnetic length is
\begin{equation}
    \frac{r_c}{\ell_B} \sim \sqrt{ \frac{T}{\hbar \omega_c}}.
\end{equation}
Hence, at high temperature the problem is effectively classical and at low temperature quantum effects can appear, as signaled by the size of the classical orbit approaching the magnetic length.

We are primarily interested in the squared displacement. At the classical level, the trajectory is
\begin{equation}
    x(t) = r_c \cos( \omega_c t ), \,\,     y(t) = r_c \sin( \omega_c t) .
\end{equation}
The squared displacement, averaged over a thermal distribution for the velocity at temperature $T = 1/\beta$, is
\begin{equation}
    \langle  (x(t)-x(0))^2 + (y(t)-y(0))^2 \rangle = \langle v^2 \rangle \frac{4}{\omega_c^2} \sin^2 \frac{\omega_c t}{2} = \frac{8}{ m \beta \omega_c^2}  \sin^2 \frac{\omega_c t}{2}.
    \label{eq:2d_Delta_c}
\end{equation}
Quantum effects modify this result when $\hbar \omega_c \gtrsim T$.

Working in Landau gauge, the full quantum result is obtained by expressing the position operator in terms of the Landau level ladder operator, $a$, and momentum in the $y$-direction, $p_y$, 
\begin{equation}
    x = \sqrt{\frac{\hbar}{2 m \omega_c}} (a + a^\dagger) + \frac{p_y}{m \omega_c}.
\end{equation}
The corresponding Heisenberg operator is 
\begin{equation}
    x(t) = \sqrt{\frac{\hbar}{2 m \omega_c}} (a e^{- i \omega_c t} + a^\dagger e^{i \omega_c t}) + \frac{p_y}{m \omega_c}.
\end{equation}
From these, we get the thermally averaged correlator,
\begin{equation}
   \langle x(t)x(0) \rangle = \frac{\hbar}{2 m \omega_c} \left( [n_B +1] e^{-i\omega_c t} + n_B e^{i \omega_c t} \right) + \sum_{k_y} \frac{\hbar^2 k_y^2}{m^2 \omega_c^2},
\end{equation}
where we note that the final term does not depend on time.

The average square displacement is 
\begin{equation}
    \langle (x(t) - x(0))^2 \rangle = 2 \langle x(0)^2 \rangle - \langle x(t) x(0) \rangle - \langle x(0) x(t) \rangle = \frac{2 \hbar}{ m \omega_c} (2 n_B + 1)\sin^2 \frac{ \omega_c t }{2}. \label{eq:2d_Delta_q}
\end{equation}
In the limit that $T \gg \hbar \omega_c$, we have $n_B \sim (\beta \hbar \omega_c)^{-1}$ and \eqref{eq:2d_Delta_q} approaches half of \eqref{eq:2d_Delta_c}; the other half comes from including the average of $(y(t)-y(0))^2$.

\subsection{Uniform Field in $N$ Dimensions}

Returning now the $N$-dimensional problem, the key new effect is that we have a distribution of the field strengths $\lambda$. The $N$-dimensional averaged square displacement is thus
\begin{equation}
    \Delta(t) = \frac{1}{N} \int^{2 \Bb}_{0} d\lambda \rho(\lambda) \frac{4 \hbar}{\lambda}(2 n_B(\lambda/m) + 1) \sin^2 \frac{\lambda t}{2m} .
\end{equation}
At zero temperature, with $n_B=0$, we can obtain a slow logarithmic increase with $t$. 

At non-zero temperature, the dominant effect comes from the classical contribution, which can be estimated as follows. We adjust the upper limit to take into account the freezing out of the classical motion for planes with a high cyclotron frequency. Then at large $t$, all but the smallest frequencies contribute an erratic amount proportional to $\sin^2  \frac{\lambda t}{2m}$. Replacing this by its value averaged over one cycle, we estimate
\begin{equation}
    \Delta(t) \sim \frac{1}{N} \int^{m T/\hbar}_{2m/t} d\lambda \rho(\lambda) \frac{4 m}{\beta \lambda^2} \sim \frac{4 m}{\beta \pi \Bb} \int^{m T/\hbar}_{2m/t} d \lambda \frac{1}{\lambda^2} = \frac{2}{\beta \pi \Bb} t + \cdots.
    \label{eq:uniform_B_diffusion}
\end{equation}
The first $\sim$ is replacing $\sin^2 \frac{\lambda t}{2m}$ with $1/2$ and putting the lower cutoff where the erratic contributions to $\Delta$ begin. The second $\sim$ is approximating the density of $\lambda$ by its central value, which is appropriate since the low field planes are the ones which contribute significantly to $\Delta$. The modes with $\lambda < 2m/t$ give a contribution of the same order. The final result shows that the particle moves diffusively at large $t$ (cutoff only when $t \sim m/\lambda_{\min} \sim N$). It is also possible to perform the integral exactly using Bessel functions.

\subsection{Imaginary Time Dynamics Within Mean Field Theory}

Strictly speaking, the mean field theory developed in section \ref{sec:EOM} doesn't apply to the case of constant field. This is because it is impossible to choose a gauge where the $A$-field has statistical translation symmetry for constant fields. This, however, is a matter of fine print. One can get arbitrarily close to constant field by choose $f(\Delta)$ to be a positive function whose slope is a constant $f'(\Delta)=-\Bb^2$ in the range of $\Delta$s we care about. In this limit, equation
 \eqref{eq:thermoEOM} for the imaginary-time dynamics becomes quadratic in the Fourier transform $G_\omega$, for all $\omega \neq 0$. We have the imaginary-time equation
 \begin{equation}
     (m\omega^2 +\Bb^2\omega^2 G_\omega)G_\omega=1.
 \end{equation}
 
 It can be solved exactly, giving us 
\begin{equation}
    G_\omega=\frac{2}{m\omega^2+\sqrt{m^2\omega^4+4\Bb^2\omega^2}}.
\end{equation}
If we are interested in large $\beta$s and small $\omega$s, we can simplify further to
\begin{equation}
    G_\omega=\frac{1}{\Bb |\omega|}
\end{equation}
 
\begin{equation}\Delta(\t) \sim \frac{2}{\pi^2 \Bb }\log \beta\sin \frac{ \pi \t}{\beta}+\mathcal O(1).
\label{eq:largeBetaConstant}
\end{equation}

\subsection{Spatially Varying Field}

Now consider the problem of a spatially varying field. With our conventions, the vector potential has units of momentum and its correlations are translation invariant and given by
\begin{equation}
    \overline{A_i(x) A_j(0)} = \delta_{ij} f(x^2/N).
\end{equation}
We will focus on the case in which
\begin{equation}
    f(u) = \frac{\hbar^2 J^2}{u + \ell^2},
\end{equation}
although other choices of $f$ can be similarly analyzed. $J$ is a dimensionless coupling which dials the strength of the local magnetic and $\ell$ is a characteristic length scale which sets the correlation length of the random vector potential. 

The magnetic field is obtained from $A_i(x)$ as
\begin{equation}
    B_{ij} = \partial_i A_j - \partial_j A_i,
\end{equation}
and its correlations can be straightforwardly obtained from those of $f$. At large $N$, the leading contribution to the magnetic field correlation is 
\begin{equation}
    \overline{\partial_i A_j(x) \partial_k A_l(0)} = - \frac{\delta_{ik} \delta_{j l}}{N} f'(x^2/N) + \cdots,
\end{equation}
which in our case is
\begin{equation}
   \overline{\partial_i A_j(x) \partial_k A_l(0)} = \frac{\delta_{ik} \delta_{j l} }{N} \frac{\hbar^2 J^2}{(x^2/N+\ell^2)^2} + \cdots.
\end{equation}

To analyze the dimensionless parameters in the problem, it is convenient to choose $\ell$ as a unit of length. Combining this length with the mass $m$ and $\hbar$ we get a characteristic energy
\begin{equation}
    E_0 = \frac{\hbar^2}{m\ell^2}.
\end{equation}
The characteristic scale of the magnetic field is
\begin{equation}
    \Bb \sim \frac{\hbar J}{\ell^2}.
\end{equation}
As discussed just above, the magnetic field is a tensor with a characteristic semi-circle spectrum of field strengths (which now vary in space), and $\Bb$ sets the overall scale of the spectrum. The corresponding magnetic length is 
\begin{equation}
    \frac{\ell_B}{\ell} \sim \frac{1}{\sqrt{J}}
\end{equation}
and the magnetic energy is 
\begin{equation}
 \frac{\hbar \omega_c}{E_0} \sim J .
\end{equation}

Putting this all together, the physics can be classified in terms of two dimensionless ratios,
\begin{equation}
    \frac{r_c}{\ell_B} \sim \sqrt{\frac{T}{\hbar \omega_c}} \sim \sqrt{ \frac{1}{J} \frac{T}{E_0}}
\end{equation}
and
\begin{equation}
    \frac{\ell_B}{\ell} \sim \frac{1}{\sqrt{J}}.
\end{equation}
In units where $\ell$, $m$, and $\hbar$ are all set to one, we see that the problem is classical for $T \gg J$ and quantum for $T \ll J$. In addition to setting the quantum-classical boundary, $J$ also plays the role of the coupling, with larger $J$ leading to larger deviations from the physics in no magnetic field. 

From the analysis in this appendix, we can form some intuition as to how the physics depends on $T$ and $J$. When $T \gg J$, all the local eigen-planes of $B_{ij}$ are effectively classical, whereas when $T \ll J$, only the low field eigen-planes remain classical. It is this smaller number of ``active'' planes which should be responsible for the low-temperature dynamics. Moreover, according to the uniform field analysis, \eqref{eq:uniform_B_diffusion}, the average square displacement after a thermal time, $t \sim \beta$, is $1/J$. Hence, the larger $J$ is, the slower the particle is moving. In the main text, we indeed find that larger $J$ leads to slower motion in many senses, including by increasing the dynamical exponent at low temperature.

\section{Analytical Solution at $J=0$}\label{app:analyticalJ0}
To simplify the formulas, we set $m=1$. We also include the confining potential as a convenient regulator and only set it to zero at the end of the calculation.

\subsection{Imaginary Time}
At limit $J\to 0$, the imaginary time Green's function in freqeuncy domain is
\begin{equation}
    G(\iu \omega_n) = \frac{1}{\omega_n^2 + \omega_0^2},
\end{equation}
where $\omega_0^2 \equiv \epsilon$. 

We can perform the Matsubara frequency summation to obtain
\begin{equation}
    \begin{split}
        G(\tau) &= \frac{1}{\beta} \sum_n G(\iu \omega_n)e^{-\iu \omega_n \tau} \\
        & = \frac{e^{-\tau  \omega_0} \left(\left(e^{2 \tau  \omega_0}+1\right) n_B(\omega_0)+1\right)}{2 \omega_0}.
    \end{split}
\end{equation}
From $\Delta(\tau) =2G(0) - 2G(\tau)$ we get
\begin{equation}
    \begin{split}
        \Delta(\tau) &= \frac{2 e^{-\beta  \omega_0} \text{csch}\left(\frac{\beta  \omega_0}{2}\right) \sinh \left(\frac{\tau  \omega_0}{2}\right) \sinh \left(\frac{1}{2} \omega_0 (\beta -\tau )\right)}{\omega_0} \\ 
        & = \frac{\tau  (\beta -\tau )}{ \beta }+ \tau  (\tau -\beta )\omega_0 + O\left(\omega_0^2\right), \\
    \end{split}
\end{equation}
at which point it is safe to take $\omega_0 \to 0$.

\subsection{Real Time}

We start with the basic formulas 
\begin{equation}
    G_R(\omega) = \frac{1}{(\omega + \iu 0^+)^2 - \omega_0^2},
\end{equation}
\begin{equation}
    \rho_G(z) = \frac{1}{2\omega_0} \delta(z-\omega_0) + \frac{1}{-2\omega_0} \delta(z+\omega_0),
\end{equation}
\begin{equation}
    G_>(t) = \frac{-\iu}{2\omega_0} n_B(\omega_0) \left( e^{-\iu \omega_0 t} e^{\beta \omega_0} - e^{\iu \omega_0 t} \right),
\end{equation}
\begin{equation}
    G_<(t) = \frac{-\iu}{2\omega_0} n_B(\omega_0)\left( e^{-\iu \omega_0 t} - e^{\iu \omega_0 t} e^{\beta \omega_0} \right).
\end{equation}
The zero-temperature version is actually given by
\begin{equation}
    \begin{split}
        G_R(t) &= \int \frac{\diff{\omega}}{2\pi} \frac{1}{2 \omega _0} \left( \frac{1}{-\omega _0+(\omega +i \eta )}-\frac{1}{\omega _0+(\omega +i \eta )} \right) \\
        &= \iu \Theta(t) \frac{e^{i t \omega _0}-e^{-i t \omega _0}}{2 \omega _0}
    \end{split}
\end{equation}

Notice that $G_R(t)$ is actually well defined in the $\omega_0 \to 0$ limit, where $G_R(t)=\iu \Theta(t) t$. But for $G_>,G_<,G_K,G_W$, the limit is not well defined.
After subtracting the appropriate zero point, $\Delta(t)$ becomes
\begin{equation}
    \Delta_>(t) = \frac{2 \operatorname{csch}\left(\frac{\beta  \omega_0}{2}\right) \sin \left(\frac{t \omega_0}{2}\right) \sin \left(\frac{1}{2} \omega_0 (t+i \beta )\right)}{\omega_0}
\end{equation}
We find that $\Re{\Delta_>(t)} > 0$. If we take $\epsilon\to 0$, we obtain the same formula via analytic continuation,
\begin{align}
    \Delta_>(t) &= \frac{t (t+i \beta )}{\beta }\label{suppeq:Deltag_real_free}\\
    \Delta_<(t) &= \frac{t (t-i \beta )}{\beta }.\label{suppeq:Deltal_real_free} 
\end{align}


\subsection{Wightman Green's Function}
The Wightman Green's function is defined by
\begin{equation}
    \iu G_W(t) \equiv \langle x(t-\iu \beta/2) x(0) \rangle .
\end{equation}
Therefore we can derive it using non-equilibrium Green's function
\begin{equation}
    \begin{split}
        \iu G_W(t) &= \iu G_>(t-\iu \beta/2) \\
        &= \iu \int \frac{\diff \omega}{2\pi} 2\pi \iu n_B(-\omega) \rho_G(\omega) e^{-\beta \omega/2} e^{-\iu \omega t}\\
        \Rightarrow G_W(\omega)& = -\iu \pi \frac{1}{\sinh(\beta \omega/2)}\rho_G(\omega) \\
        &= \frac{1}{2\sinh(\beta \omega/2)} \left(G_R(\omega) - G_R^{\dagger}(\omega) \right) .\\
    \end{split}
\end{equation}

For $\Delta$, we have $\Delta_W(\omega)= -2\iu G_W(\omega)$ for $\omega\neq 0$. The $\omega=0$ zero point needs special treatment due to the condition
\begin{equation}
        \Delta_>(t-\iu \beta/2,0) = 2\iu(G_>(t-\iu \beta/2,t-\iu \beta/2)+G_>(0,0) - 2 G_>(t-\iu \beta/2,0)). \\
\end{equation}
With translation invariance, the correlator is
\begin{equation}
    \Delta_>(t-\iu \beta/2) = 2\iu(G_>(0) - G_>(t-\iu \beta/2)),
\end{equation}
which leads to the Wightman Green's function at $t=0$
\begin{equation}
    \Delta_W(0) = 2\iu(G_>(0) - G_>(-\iu \beta/2)) = (-i)\Delta_{\tau}(\beta/2) ,\\
\end{equation}
$\Delta_{\tau}$ is imaginary time Green's function. This is explicitly true for the $J=0$ case, where from the analytic continuation we get
\begin{equation}
    \Delta_W(t) = \Delta_>(t-\iu \beta/2) = \frac{t^2 + \beta^2/4}{\beta}.
\end{equation}

\section{Other Choices of Vector Potential Correlations}
\label{app:other_f}

\begin{figure}[t]
\centering
\includegraphics[scale=0.53]{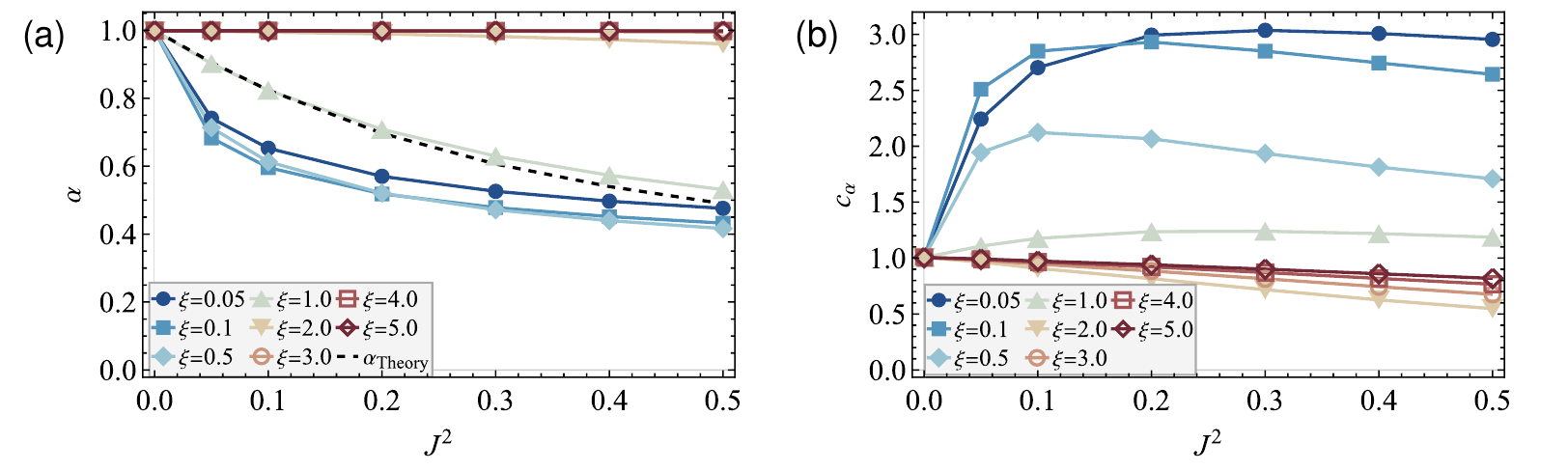}
\caption{The vector potential with different scaling law $\langle A_i(x_1) A_j(x_2)\rangle=\delta_{ij} J^2 \ell^{2(\xi-1)} / \left( \ell^2+\frac{|x_1-x_2|^2}{N}\right)^{\xi}$. The exponent $\alpha$ and coefficient $c_{\alpha}$ is defined in the Eq.~\eqref{eq:Delta_ansatz}. We extract the exponent in the imaginary time numerics with $\beta=1000$. (a) The exponent $\alpha$ for different $\xi$ and different $J^2$. We label the $\alpha_{\text{Theory}}$ in Eq.~\eqref{eq:JAlpha} as a black dashed line. (b) The coefficient $c_{\alpha}$. }
\label{fig:different_scaling}
\end{figure}

It is an interesting question to ask what's special about the $1/\Delta$ form of the vector potential covariance. To answer this question, we can generalize the vector potential correlations to
\begin{equation}
    \langle A_i(x_1) A_j(x_2)\rangle=\delta_{ij} J^2 \ell^{2(\xi-1)} / \left( \ell^2+\frac{|x_1-x_2|^2}{N}\right)^{\xi},
\end{equation}
with arbitrary parameter $\xi >0$. We can examine the behavior by performing imaginary time calculations for different $\xi$ and trying to compare to the scaling ansatz in Eq.~\eqref{eq:DeltaAnsatz}. We try to extract an effective value of $\alpha$ and coefficient $c_{\alpha}$ from $\Delta(\tau)$ using the ansatz 
\begin{equation}
  \Delta(\tau)=c_{\alpha} \left( \tau(\beta-\tau)/\beta \right)^{\alpha}  \label{eq:Delta_ansatz}
\end{equation}
In Fig.~\ref{fig:different_scaling}, we demonstrate the result for different $\xi$. Only $\xi=1$ gives a result consistent with the IR value $\alpha_{\text{Theory}}$ obeying Eq.~\ref{eq:JAlpha}. When $\xi>1$, we expect the system  will flow to the free theory, which corresponds to $\alpha=1$ and a $J$ dependent coefficient $c_{\alpha}$. This is consistent with the data in  Fig.~\ref{fig:different_scaling}.

\section{Schwinger-Keldysh Path Integral}\label{suppsec:realtime}
   \subsection{Definition of Keldysh Rotation}
   To establish the Schwinger-Keldysh formalism, we introduce the double contour with $x_{+}, x_{-}$ fields. In order to construct other useful Green's function, we perform a Keldysh rotation on the original basis and introduce the new basis $x_{cl}, x_{q}$:
   \begin{equation}\label{eq:Keldysh_rotation_x}
   	\begin{split}
   		\begin{pmatrix}
   			x_{cl} \\
   			x_{q} \\
   		\end{pmatrix} = M
   		\begin{pmatrix}
   			x_{+} \\
   			x_{-} \\
   		\end{pmatrix}
   	\end{split}
   \end{equation}
   Here the transformation matrix is $M= \frac{1}{\sqrt{2}}\begin{pmatrix}
   	1 & 1 \\
   	1 & -1 \\
   \end{pmatrix}$.
   Correspondingly, the new Green's function or self-energy is defined on such basis, which reads
   \begin{equation}\label{eq:Keldysh_rotation_G_Sigma}
   	\begin{split}
   		\quad
   		&\begin{pmatrix}
   			G_K & G_R \\
   			G_A & 0  \\
   		\end{pmatrix}
   		\equiv
   		-\iu \left<\begin{pmatrix}
   			x_{cl} \\
   			x_{q} \\
   		\end{pmatrix}
   		\begin{pmatrix}
   			x_{cl} & x_{q}\\
   		\end{pmatrix}\right> 
   		= M \begin{pmatrix}
   			G_{++} & G_{+-} \\
   			G_{-+} & G_{--}  \\
   		\end{pmatrix} M^T \\
   		& \begin{pmatrix}
   			0 & \Sigma_A \\
   			\Sigma_R & \Sigma_K  \\
   		\end{pmatrix}
   		\equiv (M^T)^{-1} \begin{pmatrix}
   			(-1)^{0+0}\Sigma_{++} & (-1)^{0+1} \Sigma_{+-} \\
   			(-1)^{1+0} \Sigma_{-+} & (-1)^{1+1}\Sigma_{--}  \\
   		\end{pmatrix} (M)^{-1}. \\
   	\end{split}
   \end{equation}
   The rationale behind defining $\Sigma_R$, $\Sigma_A$, and $\Sigma_K$ is to maintain the structure of the Schwinger-Dyson equation $(G_0^{-1} - \Sigma) \circ G = \mathbb{I}$ invariant after the Keldysh rotation.
   
   It's noticed that the statistical translational symmetry of the magnetic field still appears in our real-time formalism, even though $\delta(t_{12})$ term doesn't appear in the $\Sigma_{\gtrless}$. This is because the analog of the Schwinger-Dyson equation in real-time involves the retarded (or advanced) component of $G$ and $\Sigma$ after the Keldysh rotation. The retarded component $\Sigma_R$ should, in principle, be represented as $\Sigma_{+-}$, $\Sigma_{+-}$, $\Sigma_{++}$, $\Sigma_{++}$, where the delta function term could enter the self-energy through $\Sigma_{++}$ and $\Sigma_{--}$. 
   Similar to the imaginary time formalism, we should also expect $\Sigma_R(\omega=0)=0$ in real-time dynamics. This is consistent with the fluctuation dissipation theorem $2\Im \Sigma_R(\omega) = \tanh(\frac{\beta \omega}{2})\Sigma_K(\omega)$, which is always true in the Keldysh formalism.

   \subsection{Calculation of the Rotated Self-Energy}
    Since the effective action Eq.~\eqref{eq:FullActionReal} involves the correlator $\Delta_{ab}$, it leads to the unusual delta function term in the diagonal component of the self-energy in Eq.~\eqref{eq:selfE_ab}. We need to prove that it still leads to the correct structure of self-energy. The structure of self-energy after and before Kelydsh rotation is
    \begin{equation}
    \begin{pmatrix}
			0 & \Sigma_A \\
			\Sigma_R & \Sigma_K  \\
		\end{pmatrix}=\frac{1}{2}
	\begin{pmatrix}
			\Sigma _{++}-\Sigma _{+-}-\Sigma _{-+}+\Sigma _{--} & \Sigma _{++}+\Sigma _{+-}-\Sigma _{-+}-\Sigma _{--} \\
 \Sigma _{++}-\Sigma _{+-}+\Sigma _{-+}-\Sigma _{--} & \Sigma _{++}+\Sigma _{+-}+\Sigma _{-+}+\Sigma _{--} \\
		\end{pmatrix} 
    \end{equation}
    To begin with, we show all central results in Tab.~\ref{tab:realtime_selfE}
    \begin{table}[htb]
        \centering
        \renewcommand{\arraystretch}{1.5}
        \begin{tabular}{|c|c|}
        \hline
        Constraint & 0 = $\Sigma_{++}(t)+\Sigma_{--}(t)-\Sigma_{-+}(t)-\Sigma_{+-}(t)$ \\
        \hline
         Keldysh & $\Sigma_K(t)=\Sigma_{-+}(t)+\Sigma_{+-}(t)$ \\
           \hline
         Retarded &  $\bar{\Sigma}_R(t) \equiv  \Theta(t)\left(\Sigma_{-+}(t) -\Sigma_{+-}(t)\right)$,\ \  $
            \Sigma_R(t) =  \bar{\Sigma}_R(t) - \delta(t) \bar{\Sigma}_R(\omega=0)$  \\
            \hline
          Advanced & $\bar{\Sigma}_A(t) \equiv  \Theta(t)\left(\Sigma_{+-}(t) -\Sigma_{-+}(t)\right)$,\ \  $
            \Sigma_A(t) =  \bar{\Sigma}_A(t) - \delta(t) \bar{\Sigma}_A(\omega=0)$  \\
            \hline
        Lesser and Greater & $\Sigma_{ab}(t) = \iu \partial_{t}^2 f(\Delta_{ab}(t)) + \iu f'(\Delta_{ab}(t))\partial_{t}^2 \Delta_{ab}(t)$, $a\neq b$\\
            \hline
            Fluctuation-Dissipation Theorem & $ \Sigma_R(\omega) - \Sigma_A(\omega) = \tanh(\frac{\beta\omega}{2}) \Sigma_K(\omega) $ \\
            \hline
        \end{tabular}
        \caption{The real-time self-energy structure.}
        \label{tab:realtime_selfE}
    \end{table}
    In the following subsection, we will derive the first four results below. The lesser and greater self-energy are by definition. The fluctuation-dissipation theorem for the self-energy can be a consistent check for our special Schwinger-Keldysh structure, which requires $\Sigma_R(\omega=0) - \Sigma_A(\omega=0)=0$. 
    
    \subsubsection{Constraint on $\Sigma$}
    Before we discuss each component, we can obtain some useful relations. First, we recall the definition of time-ordered and anti-time-ordered correlator
    \begin{equation}\label{suppeq:T_order}
    	\begin{split}
    		\Delta_{++}(t_1,t_2) &= \Theta(t_{12})\Delta_{-+}(t_1,t_2) + \Theta(-t_{12})\Delta_{+-}(t_1,t_2) \\
    		\Delta_{--}(t_1,t_2) &= \Theta(t_{12})\Delta_{+-}(t_1,t_2) + \Theta(-t_{12})\Delta_{-+}(t_1,t_2). \\
    	\end{split}
    \end{equation}
    Then we calculate the summation of time-ordered and anti-time-ordered self-energy
    \begin{equation}
    	\begin{split}
    		&\Sigma_{++}+\Sigma_{--}\\
    		=& -\iu \partial_{t_1}\partial_{t_2}f(\Delta_{++}(t_1,t_2)) - \iu f'(\Delta_{++}(t_1,t_2))\partial_{t_1}\partial_{t_2}\Delta_{++}(t_1,t_2)  -\iu \partial_{t_1}\partial_{t_2}f(\Delta_{--}(t_1,t_2)) - \iu f'(\Delta_{--}(t_1,t_2))\partial_{t_1}\partial_{t_2}\Delta_{--}(t_1,t_2) \\
    		&+ \delta(t_{12}) \int \diff{t_3} \Bigg( \iu f'(\Delta_{++}(t_1,t_3))\partial_{t_1}\partial_{t_3}\Delta_{++}(t_1,t_3) + \iu f'(\Delta_{--}(t_1,t_3))\partial_{t_1}\partial_{t_3}\Delta_{--}(t_1,t_3) \\
    		& \qquad\qquad\qquad -  \iu f'(\Delta_{+-}(t_1,t_3))\partial_{t_1}\partial_{t_3}\Delta_{+-}(t_1,t_3) - \iu f'(\Delta_{-+}(t_1,t_3))\partial_{t_1}\partial_{t_3}\Delta_{-+}(t_1,t_3) \Bigg) \\
    	\end{split}
    \end{equation}
    After we expand the time-ordered and anti-time-ordered correlator using Eq.~\eqref{suppeq:T_order}, we find the delta function term is exactly canceled, and the regular self-energy terms can be simplified as $\Sigma_{+-}+\Sigma_{-+}$. Therefore it immediately leads to the first result:
    
    \begin{equation}\label{appeq:constrian}
        \frac{1}{2}\left(\Sigma_{++}+\Sigma_{--}-\Sigma_{+-}-\Sigma_{-+} \right)= 0.
    \end{equation}

    \subsubsection{Keldysh Self-Energy $\Sigma_K$}
    After using the constrain relation Eq.~\eqref{appeq:constrian}, Keldysh Green's function reads
    \begin{equation}
        \Sigma_{K}\equiv \frac{1}{2}\left(\Sigma_{++}+\Sigma_{--}+\Sigma_{+-}+\Sigma_{-+} \right) = \Sigma_{+-}+\Sigma_{-+}.
    \end{equation}

	\subsubsection{Retarded Self-Energy $\Sigma_R$}
    To deal with the retarded self-energy component, we further assume time translation invariance and simplify the formula $\Sigma_{++}-\Sigma_{--}$, which reads
    \begin{equation}
        \begin{split}
            &\Sigma_{++}-\Sigma_{--}\\
            =& \iu \partial_{t_1}^2f(\Delta_{++}(t_{12})) + \iu f'(\Delta_{++}(t_{12}))\partial_{t_1}^2\Delta_{++}(t_{12})  -\iu \partial_{t_1}^2f(\Delta_{--}(t_{12})) - \iu f'(\Delta_{--}(t_{12}))\partial_{t_1}^2\Delta_{--}(t_{12}) \\
            &+ \delta(t_{12}) \int \diff{t_3} \Bigg( -\iu f'(\Delta_{++}(t_{13}))\partial_{t_3}^2\Delta_{++}(t_{13}) + \iu f'(\Delta_{--}(t_{13}))\partial_{t_3}^2\Delta_{--}(t_{13})  \Bigg) \\
            =& \left(\bar{\Sigma}_{++} - \bar{\Sigma}_{--}\right) + \delta \Sigma_1 +  \delta \Sigma_2.\\
        \end{split}
    \end{equation}
    The result can be separated into a regular part $\bar{\Sigma}_{++} - \bar{\Sigma}_{--}$ and the $\delta \Sigma_1,\delta \Sigma_2$ part corresponding to the delta function term. 
    
    Firstly we consider $\delta \Sigma_2$, which is contributed by the explicit delta function term in the self-energy $\Sigma_{++},\Sigma_{--}$, where
    \begin{equation}
        \delta \Sigma_2(t_{12}) =\delta(t_{12})\int \diff{t_3} \Bigg(- \iu f'(\Delta_{++}(t_{13}))(\partial_{t_3}^2)\Delta_{++}(t_{13}) + \iu f'(\Delta_{--}(t_{13}))(\partial_{t_3}^2)\Delta_{--}(t_{13})  \Bigg) 
    \end{equation}
    It can be further simplified using the following tricks
    \begin{equation}
        \begin{split}
            \delta \Sigma_2(t_{12})
            &=\delta(t_{12})\int \diff{t_3} \Bigg( -\iu f'(\Delta_{++}(-t_3))\partial_{t_3}^2\Delta_{++}(-t_3) + \iu f'(\Delta_{--}(-t_3))\partial_{t_3}^2\Delta_{--}(-t_3)  \Bigg) \\
            &=\delta(t_{12})\int \diff{t_3} \Bigg( -\iu \Theta(-t_3) f'(\Delta_{-+}(-t_3))\partial_{t_3}^2\Delta_{-+}(-t_3) - \iu \Theta(t_3) f'(\Delta_{+-}(-t_3))\partial_{t_3}^2\Delta_{+-}(-t_3)  \\
            &\qquad\qquad\qquad\   +\iu\Theta(-t_3) f'(\Delta_{+-}(-t_3))\partial_{t_3}^2\Delta_{+-}(-t_3) + \iu \Theta(t_3) f'(\Delta_{-+}(-t_3))\partial_{t_3}^2\Delta_{-+}(-t_3) \Bigg)  \\
            &=-2 \delta(t_{12})\int \diff{t_3} \Theta(t_3)\Bigg( \iu  f'(\Delta_{-+}(t_3))\partial_{t_3}^2\Delta_{-+}(t_3) - \iu  f'(\Delta_{+-}(t_3))\partial_{t_3}^2\Delta_{+-}(t_3) \Bigg)  \\
        \end{split}
    \end{equation}
    In the first equation we absorb variable $t_1$ into the $t_3$ due to the integration is carried out from $-\infty \to \infty$. In the second equation, we expand the time-order and anti-time-ordered correlator. In the third equation, we want to deal with the minus sign on the $t_3$ variable. For the terms with $\Theta(-t_3)$, we can change the integration variable $t_3 \to - t_3$. Besides, for the terms with $\Theta(t_3)$, we can remove the minus sign in $\Delta$ by using the condition $\Delta_{ab}(t)=\Delta_{ba}(-t)$, which origins from the fact that displacement operator $x$ is a Hermitian operator.

    Secondly, the $\delta \Sigma_1$ denotes the delta function origins from the kinks in the time-ordered or anti-time-ordered correlator $\Delta_{++},\Delta_{--}$. This contribution only appears in the combination of $\iu \partial_{t_1}^2f(\Delta_{++}(t_{12}))-\iu \partial_{t_1}^2f(\Delta_{--}(t_{12}))$.
    By expanding $\Delta_{++},\Delta_{--}$ using Eq.~\eqref{suppeq:T_order}, we can obtain the kink contribution
    \begin{equation}
        \begin{split}
        &\iu \partial_{t_1}^2  f(\Delta_{++}(t_{12}))  - \iu \partial_{t_1}^2  f(\Delta_{--}(t_{12})) \\
            =&\iu \partial_{t_1}^2 \left( \Theta(t_{12}) f(\Delta_{-+}(t_{12}))+\Theta(-t_{12}) f(\Delta_{+-}(t_{12})) \right) - \iu \partial_{t_1}^2 \left( \Theta(t_{12}) f(\Delta_{+-}(t_{12}))+\Theta(-t_{12}) f(\Delta_{-+}(t_{12})) \right) \\
            =&  2\delta(t_{12})\left( \iu \partial_{t_{12}} f(\Delta_{-+}(t_{12}))- \iu\partial_{t_{12}} f(\Delta_{+-}(t_{12})) \right) + \sgn(t_{12}) \left(\iu \partial_{t_1}^2  f(\Delta_{-+}(t_{12}))  - \iu \partial_{t_1}^2  f(\Delta_{+-}(t_{12})) \right) \\
        \end{split}
    \end{equation}
    From the second line to the third line, we have considered both the kink contribution using $\partial_t \Theta(t) = \delta(t)$, and the regular part with a sign function. Therefore, the $\delta \Sigma_1(t_{12})$ can be defined as simplified as
    \begin{equation}
        \begin{split}
            \delta \Sigma_1(t_{12}) &= 2\delta(t_{12})\left( \iu \partial_{t_{12}} f(\Delta_{-+}(t_{12}))- \iu\partial_{t_{12}} f(\Delta_{+-}(t_{12})) \right) \\
        &= -2\delta(t_{12}) \int \diff{t_3} \left(\Theta(t_3)\iu \partial_{t_3}^2  f(\Delta_{-+}(t_3)) - \Theta(t_3)\iu \partial_{t_3}^2  f(\Delta_{+-}(t_3)) \right) \\
        \end{split}
    \end{equation}
    Finally, the regular part is
    \begin{equation}
        \begin{split}
            \bar{\Sigma}_{++}-\bar{\Sigma}_{--} &= \sgn(t_{12}) \left(\iu \partial_{t_1}^2  f(\Delta_{-+}(t_{12}))  - \iu \partial_{t_1}^2  f(\Delta_{+-}(t_{12})) + \iu  f'(\Delta_{-+}(t_{12}))\partial_{t_1}^2\Delta_{-+}(t_{12}) - \iu  f'(\Delta_{+-}(t_{12}))\partial_{t_1}^2\Delta_{+-}(t_{12})\right) \\
        &=\sgn(t_{12}) \left(\Sigma_{-+}(t_{12}) -\Sigma_{+-}(t_{12})\right), \\
        \end{split}
    \end{equation}
    which immediately leads to the regular part contribution to the retarded self-energy
    \begin{equation}
        \bar{\Sigma}_R(t_{12}) \equiv \frac{1}{2}\left(\bar{\Sigma}_{++} - \bar{\Sigma}_{--} +\Sigma _{-+} -\Sigma _{+-}\right) = \Theta(t_{12})\left(\Sigma_{-+}(t_{12}) -\Sigma_{+-}(t_{12})\right).
    \end{equation}
    This is the structure that we expect for the regular retarded self-energy term. Remarkably, the delta function contribution to the retarded self-energy can also be greatly simplified to
    \begin{equation}
        \frac{1}{2}(\delta \Sigma_1 +\delta \Sigma_2)= - \delta(t_{12}) \int \diff{t_3} \Theta(t_3) (\Sigma_{-+}-\Sigma_{+-}) = -\delta(t_{12}) \bar{\Sigma}_R(\omega=0).
    \end{equation}
  This leads to the final result for retarded self-energy:
    \begin{equation}
        \begin{split}
            \Sigma_R(t) = \frac{1}{2}(\Sigma _{++}-\Sigma _{+-}+\Sigma _{-+}-\Sigma _{--}) &= \bar{\Sigma}_R(t) + \frac{1}{2}(\delta \Sigma_1 +\delta \Sigma_2) 
            =\bar{\Sigma}_R(t) - \delta(t) \bar{\Sigma}_R(\omega=0).
        \end{split}
    \end{equation}
    
    \subsubsection{Advanced Self-Energy $\Sigma_A$}
    The advanced self-energy can be easily obtained in a similar way:
    \begin{equation}
        \begin{split}
        \bar{\Sigma}_A(t) &\equiv  \Theta(t)\left(\Sigma_{+-}(t) -\Sigma_{-+}(t)\right) \\
            \Sigma_A(t) &=  \bar{\Sigma}_A(t) - \delta(t) \bar{\Sigma}_A(\omega=0). \\
        \end{split}
    \end{equation}



\section{Numerical Methods for the Real-Time Saddle Point Equations}
\label{app:numerics}

As we discuss in the main text section~\ref{ssub:numerical_method}, due to the statistical spatial translational invariance of the system, the real-time saddle point cannot be solved self-consistently using a simple mixing method since it's numerically unstable. Instead, we choose to use a generalized gradient descent protocol with a mask function.

From equation Eq.~\eqref{eq:FullAction}, we can insert the saddle point equation and represent all self-energies via $\Delta$,
\begin{equation}\label{eq:realtime_action_onlyDelta}
    \begin{split}
        \iu S/N=& - \frac 12 \log \det 
        (G^{-1})  -  \int \diff t_1 \diff t_2  \left( \frac{1}{2} \Sigma \circ G  + \frac{1}{2} I_0(G) \right) \\
        =&  - \frac 12 \log \det 
        (G^{-1})  -  \int \diff t_1 \diff t_2  \left( \frac{1}{2} \left( G_0^{-1} \circ G - \mathbb{I}\delta(t_{12}) \right)  + \frac{1}{2} I_0(G) \right) \\
        =& - \frac 12 \log \tilde{\det} 
        (-2\iu \Delta(\omega)^{-1})  -   \left(\text{const}+ \frac{1}{2} \int \frac{\diff \omega}{2\pi} G_0(\omega)^{-1} \frac{\Delta(\omega)}{-2\iu}  + \int \diff t_1 \diff t_2 \frac{1}{2} I_0(G) \right). \\
    \end{split}
\end{equation}
The last step is to decompose $\int \diff t_1 \diff t_2 I_0(G)$ and take its derivative:
\begin{equation}
    \begin{split}
         & \frac{\delta}{\delta \Delta(\omega)} \int \diff t_1 \diff t_2  \left( \frac{1}{2} I_0(G) \right) \\
         =&  \int \diff t \frac{\delta G(t)}{\delta \Delta(\omega)} \frac{\delta}{\delta G(t)}  \int \diff t_1 \diff t_2 \left( \left[\iu \partial_{t_1}\partial_{t_2} f(\Delta_{ab}(t_{12})) \right] G(t_{12}) \right) \\
         =& \int \diff t \frac{1}{\int \diff t' \frac{\delta \Delta(t')}{\delta G(t)}e^{-\iu \omega t'}} (-\Sigma(t)) \\
         =& \int \diff t \frac{1}{\int \diff t' (-2\iu\delta(t'-t) - 2\iu \delta(t)) e^{-\iu \omega t'}} (-\Sigma(t)) \\
         =& \int \diff t \frac{1}{-2\iu e^{-\iu \omega t}- 2\iu \delta(t)\delta(\omega)} (-\Sigma(t)). \\
    \end{split}
    \label{eq:realtime_action_onlyDelta_derivative}
\end{equation}

Since we're working in the $\omega\neq 0$ sector, the $\delta(\omega)$ term can be ignored. Summing all results we arrive at
\begin{equation}
    \begin{split}
        \iu \frac{\diff S}{\diff \Delta(\omega)}/N  
        =&  \frac 12 \Delta^{-1}(\omega)  +  \left( -\frac{1}{2} \left(  \Delta_0^{-1} \right) + \frac{1}{2} \frac{\Sigma(\omega)}{-2\iu} \right) \\
        =&  \frac{\iu}{4}  \left( (-2\iu)\Delta^{-1}(\omega) - (G_0^{-1} - \Sigma)  \right). \\
    \end{split}
    \label{eq:realtime_action_onlyDelta}
\end{equation}
If we only consider the retarded component, we will find the gradient to be zero is the same as the Schwinger-Dyson equation\eqref{eq:SD_R_Delta}. In the numerics, we choose the update rule to be
\begin{equation}
    \begin{split}
        \Delta(\omega)_{\text{new}} = \Delta(\omega)_{\text{old}} + \zeta \frac{\iu}{4}  \left( (-2\iu)\Delta^{-1}_{R}(\omega) - (G_{0,R}^{-1}(\omega) - \Sigma_R(\omega))  \right) \Delta_R(\omega)^2.
    \end{split}
\end{equation}
Since the gradient can diverge near small $\omega$, we introduce an extra $\Delta_R(\omega)^2$ factor as a mask function. This numerical trick stabilizes the iteration process. We consider convergence to be achieved when $||\Delta_{R,\text{new}}(\omega)-\Delta_{R,\text{old}}(\omega)|| < 10^{-6}$.


\bibliography{ref.bib}
\bibliographystyle{JHEP}

\end{document}